\documentclass[a4paper,12pt]{article}
\usepackage{graphicx}
\usepackage{float}

\title{Circular geodesics of Bardeen and Ayon-Beato-Garcia regular black-hole and no-horizon spacetimes}

\author{Zden\v{e}k Stuchl\'{i}k$^1$ and Jan Schee$^2$\\
		\emph{Institute of Physics, Faculty of Philosophy and Science,}\\
		\emph{Silesian university in Opava}\\
		\emph{Bezrucovo nam. 13, CZ-746 01 Opava, Czech Republic}\\
		e-mail: $^1$zdenek.stuchlik@fpf.slu.cz, $^2$jan.schee@fpf.slu.cz}

\date{}

\newcommand{\diff}{\mathrm{d}}
\newcommand{\beq}{\begin{equation}}
\newcommand{\eeq}{\end{equation}}
\newcommand{\bea}{\begin{eqnarray}}
\newcommand{\eea}{\end{eqnarray}}

\begin{document}
\maketitle

\begin{abstract}
We study circular geodesic motion of test particles and photons in the Bardeen and Ayon-Beato-Garcia (ABG) geometry describing spherically symmetric regular black-hole or no-horizon spacetimes. While the Bardeen geometry is not exact solution of Einstein's equations, the ABG spacetime is related to self-gravitating charged sources governed by Einstein's gravity and non-linear electrodynamics. They both are characterized by the mass parameter $m$  and the charge parameter $g$. We demonstrate that in similarity to the Reissner-Nordstrom (RN) naked singularity spacetimes an antigravity static sphere should exist in all the no-horizon Bardeen and ABG solutions that can be sorrounded by a Keplerian accretion disc. However, contrary to the RN naked singularity spacetimes, the ABG no-horizon spacetimes with parameter $g/m > 2$ can contain also an additional inner Keplerian disc hidden under the static antigravity sphere. Properties of the geodesic structure are reflected by simple observationally relevant optical phenomena. We give silhouette of the regular black hole and no-horizon spacetimes, and profiled spectral lines generated by Keplerian rings radiating at a fixed frequency and located in strong gravity region at or nearby the marginally stable circular geodesics. We demonstrate that the profiled spectral lines related to the regular black holes are qualitatively similar to those of the Schwarzschild black holes, giving only small quantitative differences. On the other hand, the regular no-horizon spacetimes give clear qualitative signatures of their presence while compared to the Schwarschild spacetimes. Moreover, it is possible to distinguish the Bardeen and ABG no-horizon spacetimes, if the inclination angle to the observer is known. 
\end{abstract}

\section*{Introduction}
Recently, black-hole solutions of alternative theories to Einstein's gravity are studied extensively to find astrophysically and observationally relevant signatures of these alternatives. One family of the alternative theories belongs to the higher-dimensional spacetimes related to String theory, as the braneworld models and their black-hole solutions restricted to the 3D brane \cite{Dadhich-etal:2000:PhLettB,Ali-Gum:2005:PHYSR4:,Sche-Stu:2008:IJMPD:,Sche-Stu:2008:GenRelGrav:}. A 3D approach to quantum gravity, based on methods of solid state physics, is developed in the framework of Ho\v{r}ava gravity with Lorentz invariance breaking \cite{Hor:2009:PHYSR4:,Hor:2009:PHYSRL:,Hor-MelT:2010:PHYSR4:,Gri-Hor-Mel:2013:PHYSRL:}. Here we focus our attention to the solutions of the 3D Einstein gravity combined with non-linear electrodynamic laws enabling avoidance of the physical singularity in the black-hole solutions \cite{AyB-Gar:2000:PhysLetB:}. 

Black holes governed by the standard general relativity contain a physical singularity with diverging Riemann tensor components and predictability breakdown, considered as a realm of quantum gravity overcoming this internal defect of general relativity. However, families of regular black hole solutions elimination the physical singularity from the spacetimes having an event horizon have been found. Of course, these are not vacuum solutions of the Einstein gravitational equations, but contain necessarily a properly chosen additional field, or modified gravity, and the energy conditions related to the existence of physical singularities \cite{Haw-Elli:1973:LargeScaleStructure:} are then violated. The first, \emph{non-exact} regular black hole solution containing a magnetic charge as a source parameter has been proposed by Bardeen \cite{Bar:1968:GR5Tbilisi:}. Ay\'{o}n-Beato and Garc\'{i}a have shown that the magnetic charge is related to a non-linear electrodynamics \cite{AyB-Gar:2000:PhysLetB:}. The \emph{exact} regular black hole solution relating the Einstein equations and a non-linear electrodynamics has been introduced by Ayon-Beato and Garcia \cite{AyB-Gar:1998:PhysRevLet:,AyB-Gar:1999:PhysLetB:,AyB-Gar:1999:GenRelGrav:}. Another approach to the regular black holes has been applied by Hayward \cite{Hay:2006:PhysRevLet:}. Modification of the mass function in the Bardeen and Hayward regular black hole solutions and inclusion of the cosmological constant has been introduced in the new solutions of Neves and Saa \cite{Nev-Saa:2014:arXiv:1402.2694:}. Rotating regular black hole solutions has been introduced in \cite{Mod-Nic:2010:PHYSR4:,Bam-Mod:2013:PhysLet:,Tos-Abd-Ahm-Stu:2014:PHYSR4:}. 

Properties of geodesic motion in the field of regular black holes have been recently discussed in several papers \cite{Gar-Hac-Kun-Lam:2013:arXiv:1306.2549:,Bam-Mod:2013:PhysLet:,Tos-Abd-Ahm-Stu:2014:PHYSR4:}. Of course, it is of crucial interest to study observationaly relevant properties of the regular black hole spacetimes and compare them with those related to the standard black hole spacetimes, or to those related to the black hole solutions inspired by the string theory, e.g., the braneworld black holes. There are three observationally important tests of the character of strong gravity in vicinity of the black hole event horizon, namely, the spectral continuum \cite{McCli-etal:2011:CLAQG:,Rem-McCli:2006:ARAA:,Fra-McCli:2014:ArXiv:}, the profiled spectral lines \cite{Laor:1989:IUAS:,Bao-Stu:1992:ApJ:,Bao-Had-Oest:1996:ApJ:,Fan-Cal-Cad:1997:,Sche-Stu:2008:GenRelGrav:}, and high-frequency quasiperiodic oscillations \cite{Tor-Abr-Klu-Stu:2005:ASTRA:,Stu-Kot:2009:GenRelGrav:}. These phenomena are governed by the geodesic structure of the spacetime, especially the circular geodesic motion is important. 

Here we concentrate our attention to the profiled spectral lines created at the innermost parts of the Keplerian (geodesic) accretion discs orbiting the regular black holes related to the Ayon-Beato-Garcia (ABG) solutions of mixed Einstein's and non-linear electrodynamic equations and compare them to those related to the Bardeen spacetimes. The profiled lines generated at these spherically symmetric regular black hole spacetimes are compared to those generated under corresponding conditions around the standard Schwarzschild black holes. 

We extend here the study of the geodesic structure to the Bardeen and ABG spacetimes with parameters adjusted in such a way that the spacetimes do not contain an event horizon. As we shall see, such "no-horizon" solutions resemble in many aspects the Reissner-Nordstrom naked singularity solutions (and partly the axisymmetric Kerr naked singularity solutions) of the standard general relativity, or the Kehagias-Sfetsos naked singularity solutions of the modified Ho\v{r}ava quantum gravity. However, in the Bardeen and ABG "no-horizon" spacetimes no spacetime singularity occurs, as components of the Riemann tensor of these spacetimes are finite everywhere, contrary to the Reissner-Nordstrom, Kerr, or Kehagias-Sfetsos spacetimes. Of course, we have to expect that at the centre of coordinates, $r=0$, the source of the electromagnetic field of the background is located and we expect a special behavior related to the electrodynamic part of the theory behind the Bardeen and ABG solutions \cite{AyB-Gar:1999:PhysLetB:, AyB-Gar:1999:GenRelGrav:}. Here we assume that at the centre $r=0$, the trajectories of test particles and photons terminate, similarly to the case of the central singular points in the spherically symmetric naked singularity spacetimes. 

We demonstrate properties of the geodesic structure of the Bardeen and ABG spacetimes giving silhouette of the black hole and no-horizon spacetimes, and calculating profiles of spectral lines generated by Keplerian (geodesic) rings radiating at a fixed frequency corresponding, e.g., to a fluorescent iron spectral line. The profiled spectral lines generated in the fully regular, "no-horizon" Bardeen and ABG spacetimes demonstrate clear distinctions while compared to those generated in the Kehagias-Sfetsos naked singularity spacetimes \cite{Stu-Sche:2014:CLAQG:,Vie-etal:2014:PHYSR4:,Stu-Sche-Abd:2014:PHYSR4:} of the Ho\v{r}ava gravity \cite{Hor:2009:PHYSR4:,Keh-Sfe:2009:PhysLetB:}, or to those generated by the rings orbiting Kerr naked singularities \cite{Stu-Sche:2010:CLAQG:,Sche-Stu:2013:JCAP:}. 

\section{Spherically symmetric regular spacetimes}

First, let us review basic properties of the spherically symmetric regular Bardeen and ABG spacetimes for both the black-hole and no-horizon families. The properties of the electromagnetic field related to the regular black hole (or no-horizon) spacetimes are discussed in \cite{AyB-Gar:1998:PhysRevLet:,AyB-Gar:1999:PhysLetB:,AyB-Gar:1999:GenRelGrav:}. We concentrate here on the geometry properties since only the Keplerian geodesic motion is considered for accretion discs and rings and optical phenomena are given by photons following the null geodesics of the spacetimes. The geometry of the Bardeen and ABG black hole (no-horizon) spacetimes is characterized in the standard spherical coordinates and the geometric units (c=G=1) by the line element
\beq
	\diff s^2=-f(r)\diff t^2+\frac{1}{f(r)}\diff r^2+r^2(\diff\theta^2+\sin^2\theta\diff\phi^2),\label{interval}
\eeq
where the "lapse" $f(r)$ function depends only on the radial coordinate. Both Bardeen and ABG spacetimes are constructed to be regular everywhere, i.e., the components of the Riemann tensor, and the Ricci scalar are finite at all $r\ge 0$ \cite{AyB-Gar:1999:GenRelGrav:}.

The lapse function $f(r)$ is given by the formulae 
\begin{enumerate}
\item Bardeen spacetime
\beq
	f(r)=1 - \frac{2 m r^2}{(g^2 + r^2)^{3/2}} , 
\eeq
\item ABG spacetime
\beq
	f(r)=	1  - \frac{2 m r^2}{(g^2 + r^2)^{3/2}} + \frac{g^2 r^2}{(g^2 + r^2)^2}.
\eeq
\end{enumerate}
Here $m$ denotes the standard gravitational mass parameter, while $g$ corresponds to the charge parameter measured in units of $m$. From here on we put in our calculations $m=1$ this means that both charge $g$ and radial coordinate $r$ are expressed in units of $m$.  

The pseudosingularities of the Bardeen and ABG spacetimes are given by the condition  
\beq
f(r)=0. \label{pseudosingularity}
\eeq 
From the definition of the coordinate time $t$ it follows that the condition (\ref{pseudosingularity}) determines the event horizons of the spacetimes. If real and positive solutions of Eq. (\ref{pseudosingularity}) exist, the spacetime governs a black hole, if there is no such solution, the spacetime is fully regular, having no event horizon, and we call it "no-horizon" spacetime. The loci of the black hole horizons are implicitly given by the relations  
\begin{enumerate}
\item Bardeen 
	\beq
		g^6 + (3g^2 - 4)r^4 + 3g^4r^2 + r^6 = 0 , 
	\eeq
\item ABG 
	\beq
	    (g^2 + r^2)^2(g^4 + 4g^2r^2 + r^4) - 4r^4(g^2 + r^2) + g^4r^4 = 0 .
	\eeq
\end{enumerate} 
The condition for loci of the event horizons of the Bardeen spacetimes is cubic in both the radius squared $r^2$ and the charge parameter $g^2$, while it is quartic relation for both squared radius and charge in the ABG spacetimes. The solutions for these loci are presented in Fig. 1.

The critical value of the parameter $g$, separating the black hole and the "no-horizon" Bardeen and ABG spacetimes reads
\begin{enumerate}
\item Bardeen 
	\beq
		g_{NoH/B} = 0.7698 ;
	\eeq
\item ABG 
	\beq
	    g_{NoH/ABG} = 0.6342 .
	\eeq
\end{enumerate} 
In the "no horizon" Bardeen and ABG spacetimes, the metric is regular at all radii $r \geq 0$, but we have to consider $r=0$ (or its small vicinity, at least of the Planck length) to be the site of the source charge of the spacetime assuming that all particle or photon trajectories reaching $r=0$ terminate there. Detailed discussion of the properties of the electromagnetic field related to the Bardeen and ABG spacetimes can be found in \cite{AyB-Gar:1999:GenRelGrav:}, however, it is not necessary for the purposes of our study. 

\section{Test particle motion}

Motion of test particles and photons is governed by the geodesic structure of the spacetime. The geodesic equations are significantly simplified in the static and spherically symmetric spacetimes due to their symmetries and can be well governed by an effective potential. The motion is always confined to a central plane. 
 
\subsection{Equations of motion and effective potential}
Both the considered regular spacetimes posses two Killing vector fields, $\partial/\partial t$ and $\partial/\partial\phi$,  implying existence of two constants of motion, energy $E\equiv -p_t$ and axial angular momentum $L\equiv p_\phi$. There is also an additional constant of motion $Q$, "latitudinal" angular momentum defined by the relation $p_\theta^2\equiv Q-L^2\cot^2\theta$. For massive particles with rest mass $\mu > 0$, the constants of motion can be related to the constant rest mass (energy) $\mu$, and are then specific energy and specific angular momenta of the motion. The equations of the geodesic motion can be written in the integrated form, giving components of the four-velocity (four-momentum) of the particle that read 
\bea
	p^t&=&\frac{E}{f(r)},\\
	(p^r)^2 &=&E^2-f(r)\left(\kappa+\frac{L^2+Q}{r^2}\right),\\
	(p^\theta)^2&=&\frac{1}{r^4}\left(Q-L^2\cot^2\theta\right),\\
	p^\phi&=&\frac{L}{r^2\sin^2\theta}
\eea
where $\kappa=0$ for photons and $\kappa=1$ for massive particles. For further analysis it is convenient to define an effective potential of the motion by the relation 
\beq
	V_{eff}=f(r)\left(\kappa+\frac{L^2+Q}{r^2}\right).
\eeq

\subsection{Circular geodesics}
\begin{figure}[ht]
\begin{center}
\begin{tabular}{cc}
\includegraphics[scale=0.7]{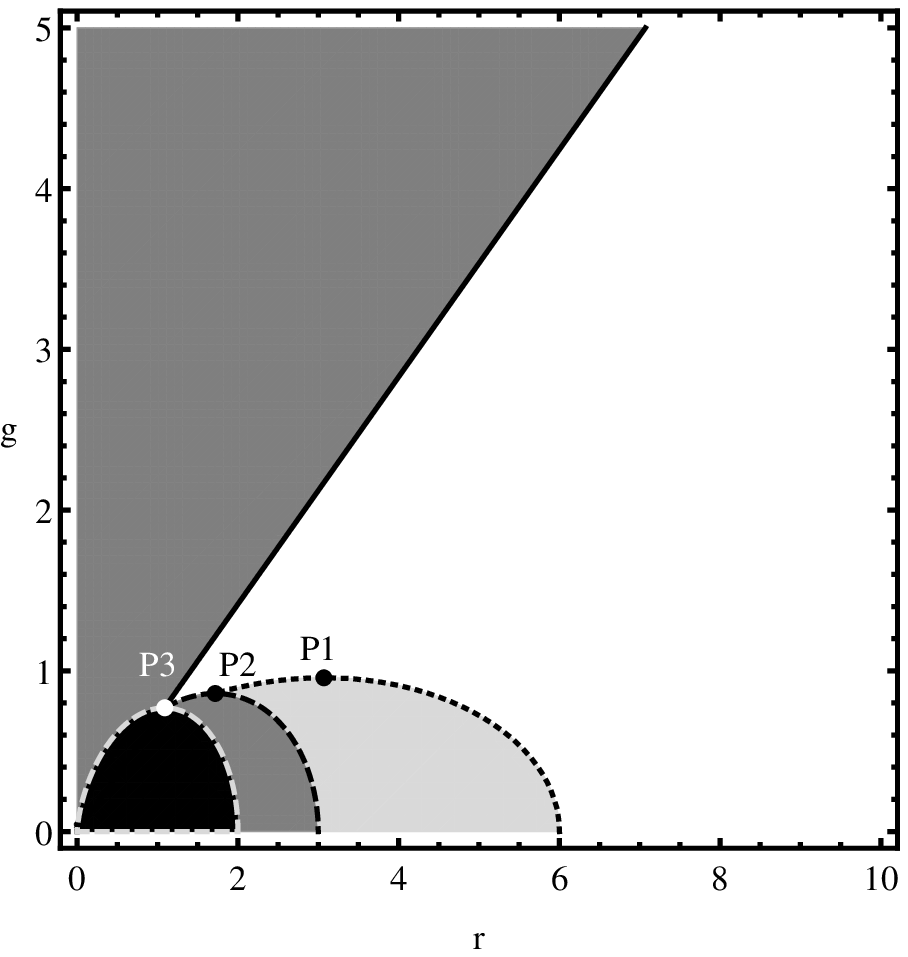}&\includegraphics[scale=0.7]{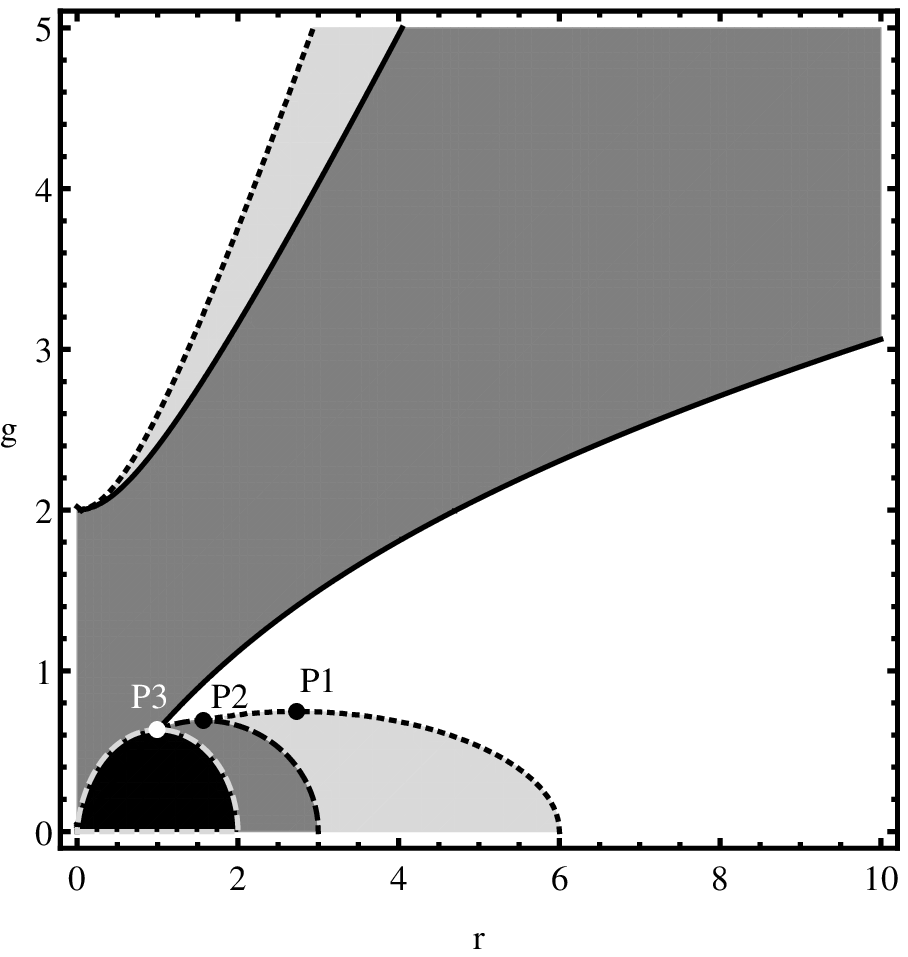}
\end{tabular}
\caption{Loci of circular geodesic orbits in the Bardeen (left) and ABG (right) spacetimes. The white regions correspond to stable orbits, while the light gray regions represent unstable orbits. In the dark gray regions no circular geodesics exist and the black region represents dynamical region of black hole.
The points on the \emph{black-dotted} curve satisfy equation  $\diff^2 V_{eff}(r,L_K,g)/\diff r^2=0$, while points on the \emph{black-dashed} curve satisfy equation $L^2_K(r,g)\rightarrow \infty$, and points on the black solid curve satisfy equation $L^2_K(r,g)=0$. We define three characteristic points (see text). For the Bardeen spacetimes, the characteristic points are $P1=(3.07,0.95629)$, $P2=(1.7179, 0.85865)$, and $P3=(1.09,0.76988)$. For the ABG spacetimes, the characteristic points are $P1=(2.73,0.74684)$, $P2=(1.57, 0.690771)$, and $P3=(0.9941,0.6343)$.  \label{fig1}}
\end{center}
\end{figure}
In the case of spherically symmetric spacetimes the motion is planar taking place in central planes which are all equivalent. For convenience one usually picks up the equatorial plane determined by equation $\theta=\pi/2$ where the constant $Q=0$.  Circular geodesics are determined by the simultaneously conditions 
\beq
	p^r=0,  \frac{\diff p^r}{\diff w}=0 \label{circ_geo_cond}
\eeq 
where $w$ is an affine parameter of the geodesic motion and in the case of massive particles it is related to their proper time $\tau$. From the conditions (\ref{circ_geo_cond}) and the equations of motions, it follows that circular geodesics in the equatorial plane must fulfil the equations
\beq
	E^2_C=V_{eff}(r_c;L_c),\quad \frac{\diff V_{eff}}{\diff r}|_c=0.
\eeq
This implies
\beq
	L^2_c=\frac{r^2 f'(r)}{2f(r)/r-f'(r)}, 
\eeq
where the prime means derivative with respect to the radial coordinate. The corresponding energy of the circular orbit follows from the condition $(p^r)^2=E^2_c-V=0$. We obtain 
\beq
	E_c^2=f(r)\left(1+\frac{L_c^2}{r^2}\right)=\frac{2f^2}{2f-r f'}.
\eeq
The angular frequency of the circular geodesics (Keplerian angular frequency) relative to distant observers reads
\beq
	\Omega_c \equiv \frac{d\phi}{dt} = \frac{f}{r^2}\frac{L_c}{E_c}.
\eeq

In the Bardeen and ABG spacetimes, the specific angular momentum $L_c$, the specific covariant energy $E_c$, and the angular frequency $\Omega_c$ of the circular orbits read
\begin{enumerate}
\item \emph{Bardeen}
\bea
	L^2_c&=&\frac{r^4 (-2 g^2 + r^2)}{-3 r^4 + (r^2 + g^2)^{5/2}},\\
	E^2_c&=&\frac{\left[-2 r^2 + (r^2 + g^2)^{3/2}\right]^2}{g^6 + 3 g^4 r^2 + 3 g^2 r^4 + r^6 - 
	 3  r^4 \sqrt{r^2 + g^2}},\\
	\Omega_c^2&=&\frac{(r^2-2g^2)}{(r^2+g^2)^{5/2}} .
\eea 

\item \emph{ABG}
\bea
	L^2_c&=&\frac{(r^4 (g^2 (g^2 - r^2) \sqrt{g^2 + r^2} + 
	    (-2 g^4 - g^2 r^2 + r^4)))}{(-3  r^4 (g^2 + r^2) + 
	 \sqrt{g^2 + r^2} (g^6 + 3 g^4 r^2 + 5 g^2 r^4 + r^6))},\\
	E^2_c&=&\frac{(g^4 \sqrt{g^2 + r^2} + r^4 (-2  + \sqrt{g^2 + r^2}) + 
	  g^2 r^2 (-2  + 3 \sqrt{g^2 + r^2}))^2}{((g^2 + r^2)^2 (g^6 + 
	   3 g^4 r^2 + 5 g^2 r^4 + r^6 - 3  r^4 \sqrt{g^2 + r^2}))},\\
	\Omega_c^2&=&\frac{ r^4 + g^4 (-2  + \sqrt{r^2 + g^2}) - g^2 r^2 ( + \sqrt{g^2 + r^2})}{(r^2+g^2)^{7/2}}.   
\eea 
\end{enumerate} 

We shall discuss the standard properties of the circular geodesics, as their stability to the radial perturbations (the motion is always stable relative to the latitudinal perturbations), the existence of photon circular orbits; in the case of the no-horizon spacetimes we determine the static radii governing so called antigravity sphere \cite{Vie-etal:2014:PHYSR4:}, and the radius of vanishing of the angular frequency gradient. The Bardeen and ABG spacetimes can be classified by the properties of the circular geodesics as demonstrated in Fig. \ref{fig1} where all the characteristic radii, including those of the black hole horizons, are illustrated in dependence on the dimensionless parameter $g$. 

In the Fig.\ref{fig1} we have introduced three characteristic parameters $g_{P1}$:
\begin{itemize}
\item Point $P_1$ is the maximum of curve representing marginaly stable orbits. For values of $g<g_{P1}$ there exist at least one marginally stable circular orbit.
\item Point $P_2$ is the local maximum of the curve given implicitly by $L^2_K(r,g)=0$. Between points $g_{P1}$ and $g_{P2}$ there exist two maginally stable orbits. 
\item Point $P_3$ is the local maximum of the curve implicitly given by $f(r,g)=0$.
\end{itemize}

\begin{figure}
\begin{center}
\begin{tabular}{cc}
\includegraphics[scale=0.6]{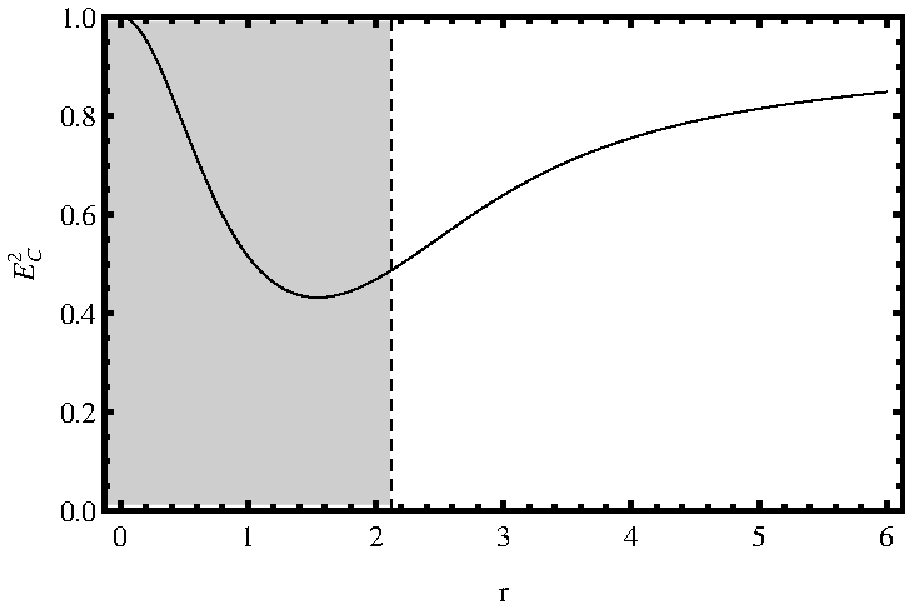}&\includegraphics[scale=0.6]{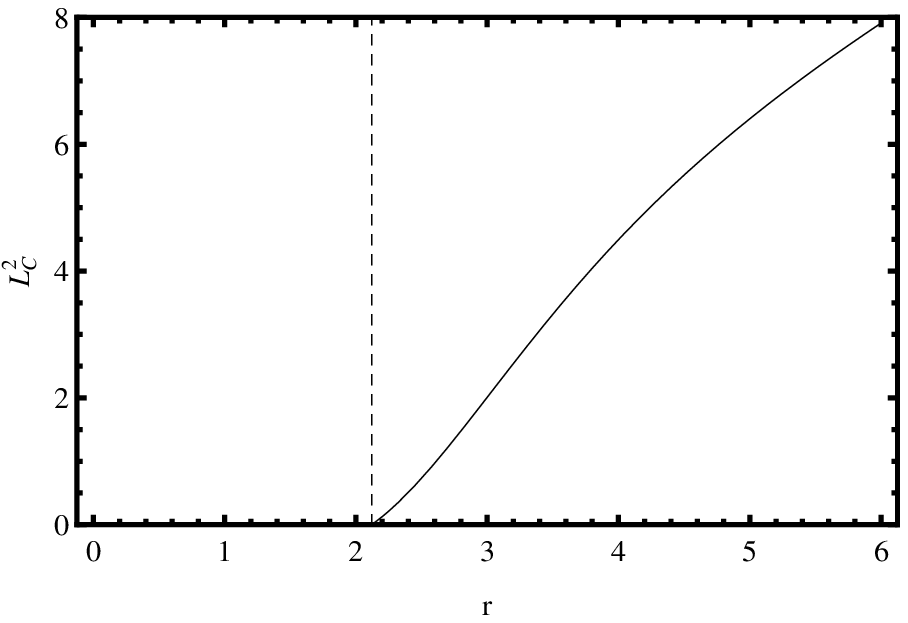}\\
\includegraphics[scale=0.6]{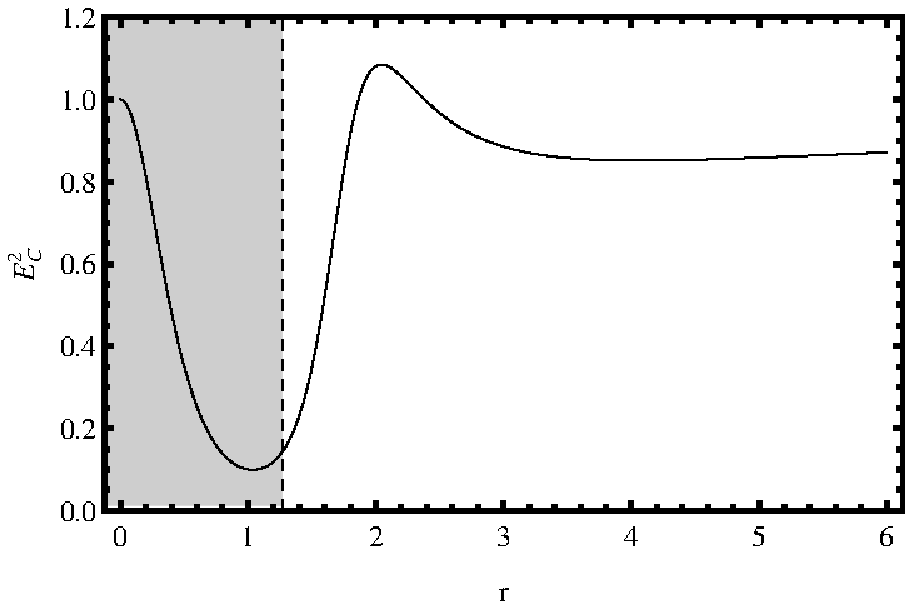}&\includegraphics[scale=0.6]{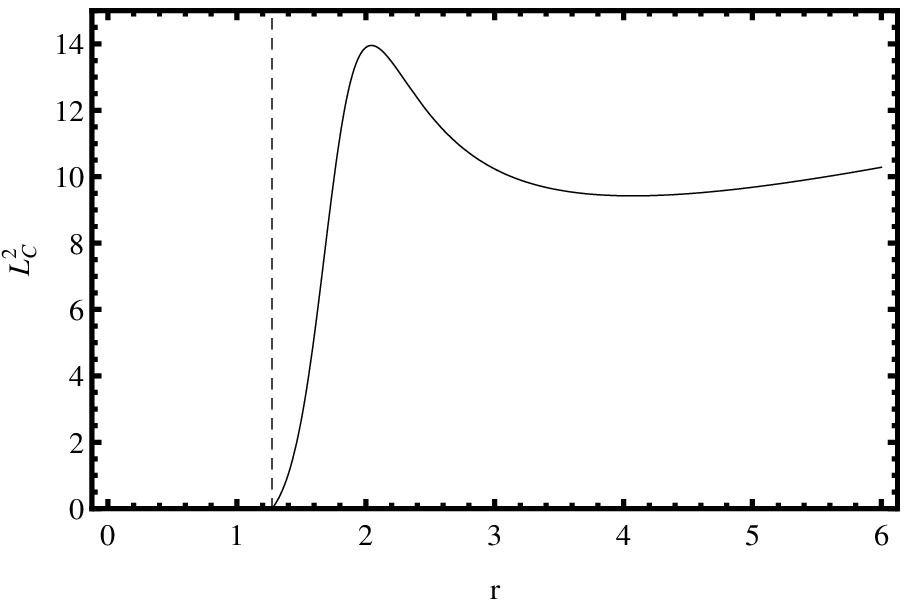}\\
\includegraphics[scale=0.6]{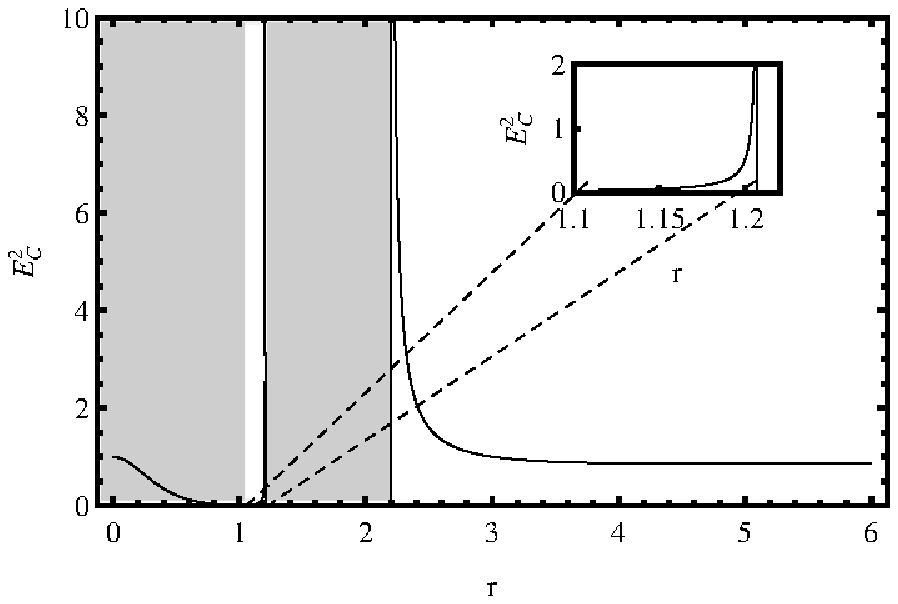}&\includegraphics[scale=0.6]{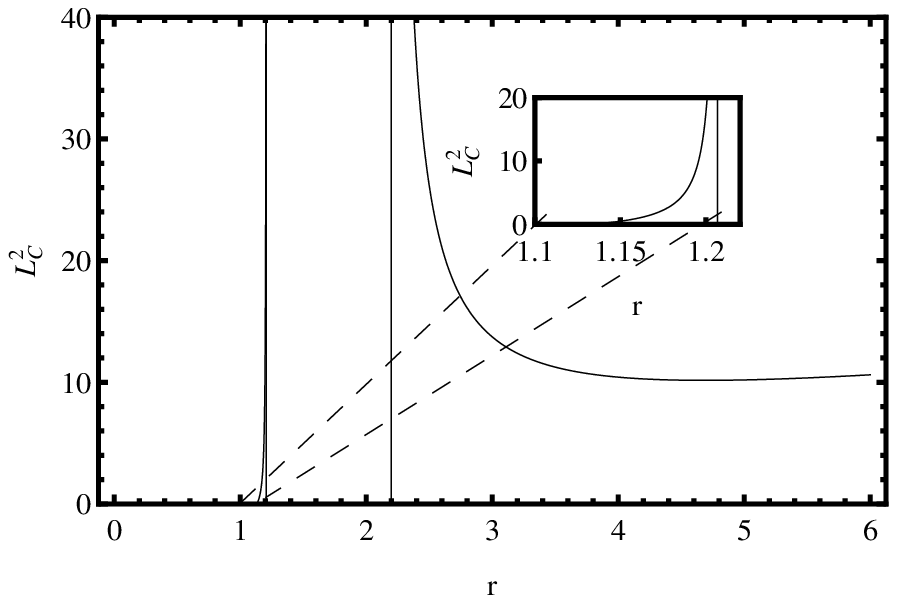}\\
\includegraphics[scale=0.6]{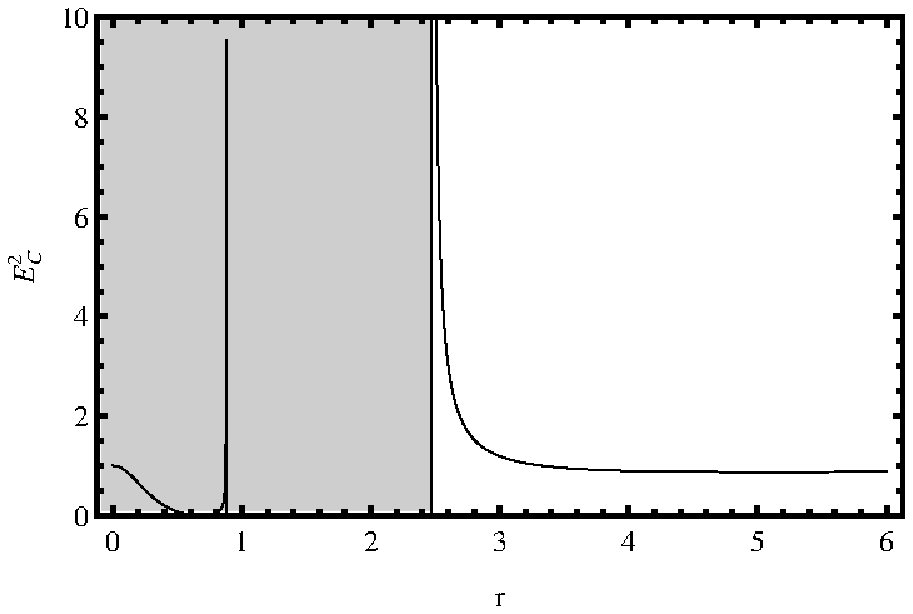}&\includegraphics[scale=0.6]{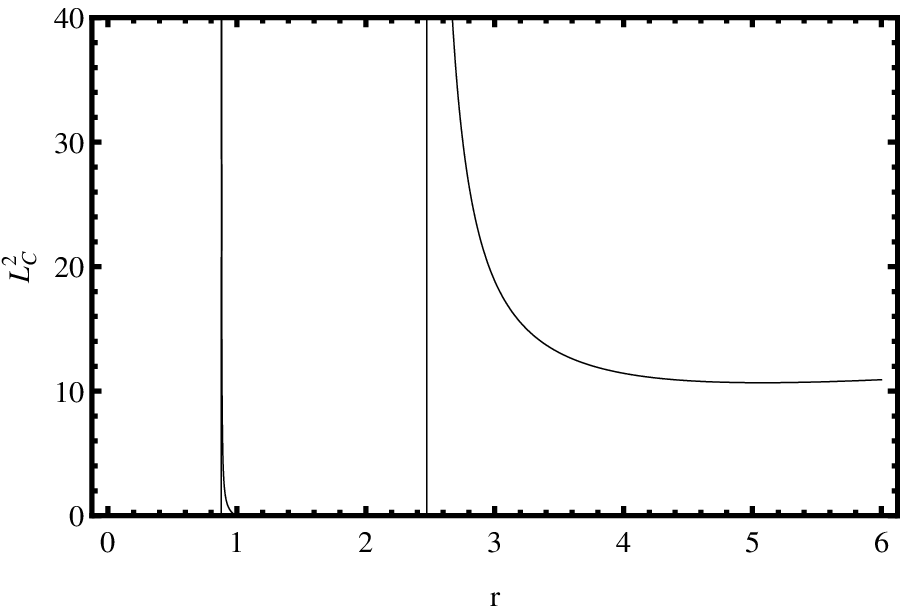}
\end{tabular}
\caption{Radial profiles of $E_C^2$ (left) and $L_C^2$ (right) given for the Bardeen spacetimes with $g=1.5 > g_{S/B}$ (top), $g_{S/B}<g=0.9<g_{P/B}$, $g_{P/B}<g=0.8<g_{NoH/B}$, and $g=0.7<g_{NoH/B}$ (bottom). Shaded are the regions where the circular geodesics are not allowed. For the ABG spacetimes with $g < 2$ qualitatively the same radial profiles occur. \label{fig.2}}
\end{center}
\end{figure}

\begin{figure}
\begin{center}
\begin{tabular}{cc}
\includegraphics[scale=0.6]{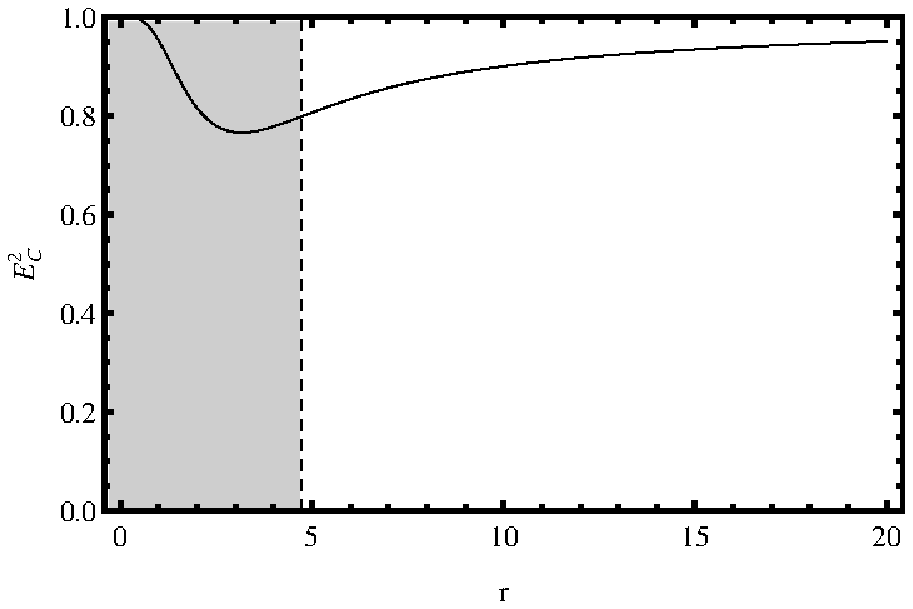}&\includegraphics[scale=0.6]{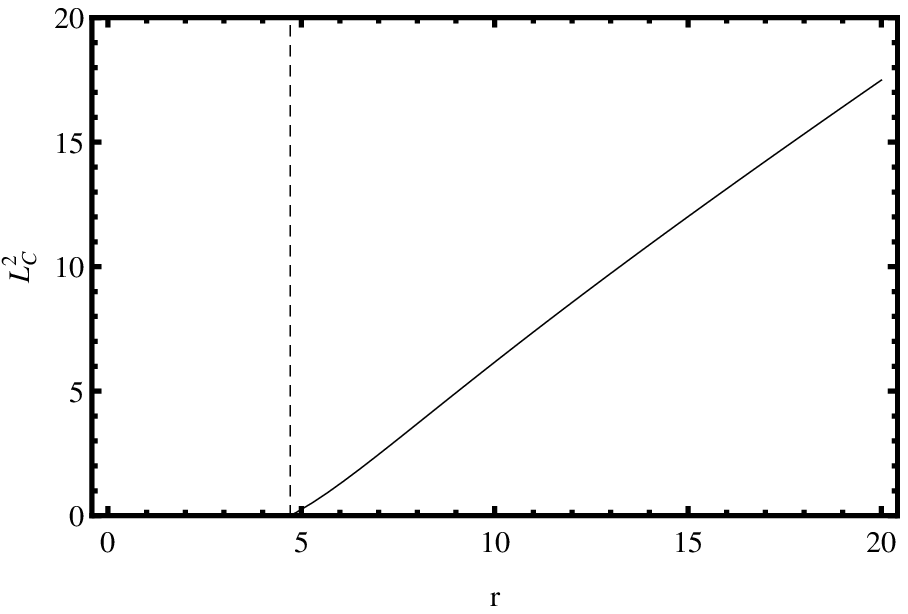}\\
\includegraphics[scale=0.6]{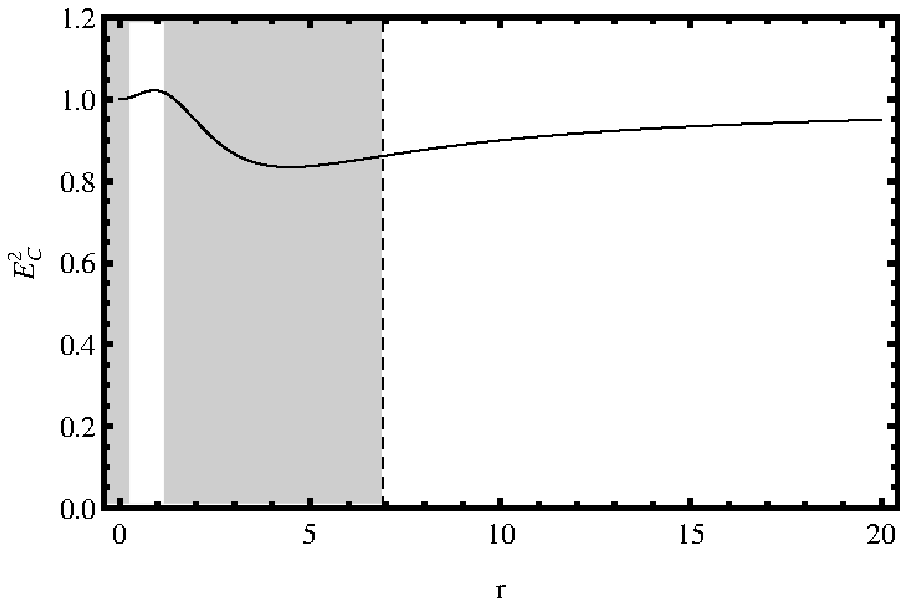}&\includegraphics[scale=0.6]{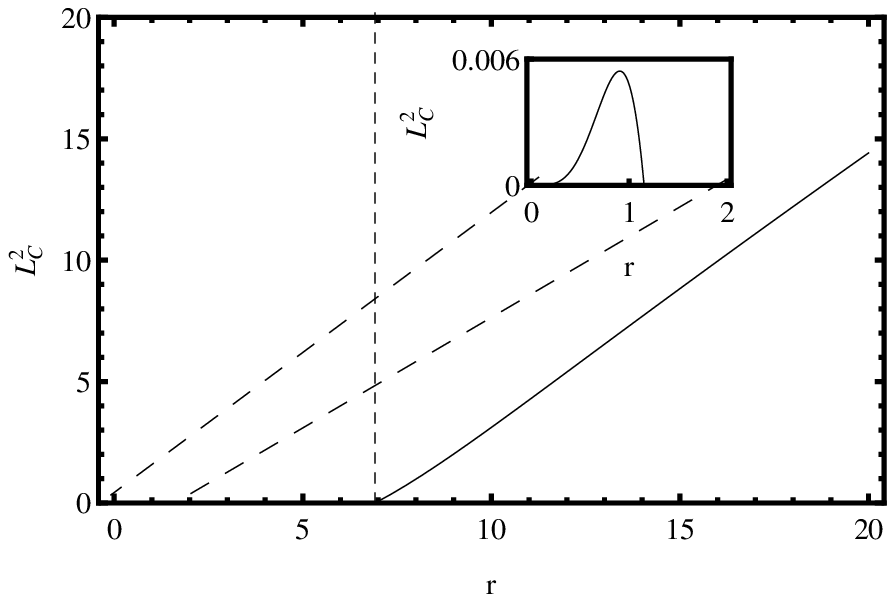}\\
\includegraphics[scale=0.6]{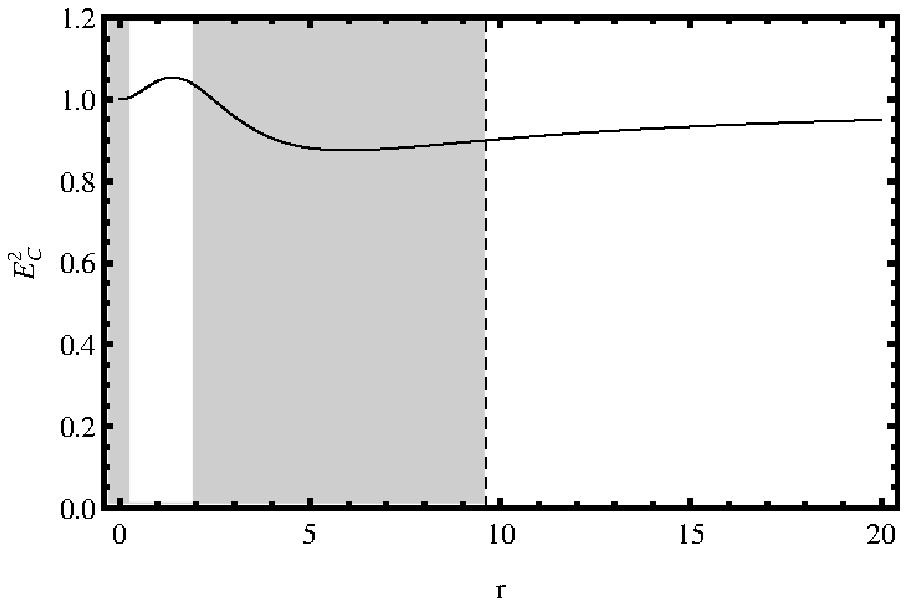}&\includegraphics[scale=0.6]{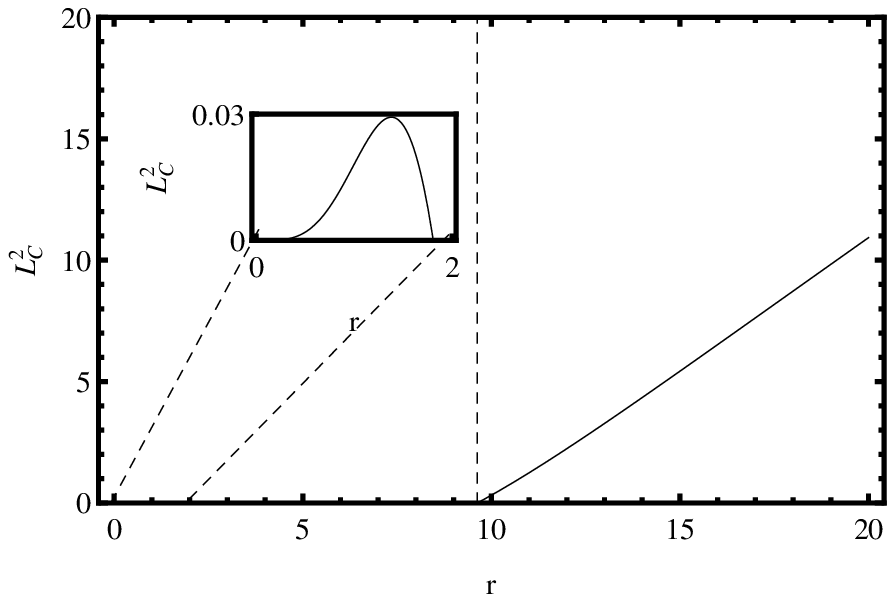}
\end{tabular}
\caption{Radial profiles of $E_C^2$ (left) and $L_C^2$ (right) are given for the special case of the ABG no-horizon spacetimes having inner Keplerian discs located under the stable static radius, with values of the charge parameter $g=2.0$ (top), $g=2.5$ (middle), and $g=3.0$ (bottom). \label{fig.3}}
\end{center}
\end{figure}

For characteristic values of the parameter $g$, given by the classification of the Bardeen spacetimes according to the properties of circular geodesics, we illustrate in Fig. \ref{fig.2} the characteristic behavior of the radial profiles of the specific energy and specific angular momentum, $E^2_{c}(r;g)$ and $L^2_{c}(r;g)$. It should be stressed that in the regions where the energy and angular momentum decrease (increase) with decreasing radius, the circular orbits are stable (unstable) relative to the radial perturbations. For the ABG spacetimes with $g < 2$, behavior of the $E^2_{c}(r;g)$ and $L^2_{c}(r;g)$ profiles is qualitatively the same as for the Bardeen spacetimes. For the ABG spacetimes with $g > 2$, an additional inner Keplerian disc occurs under the antigravity sphere, as illustrated in Fig. \ref{fig.3}. 

\subsection{Antigravity sphere}

We should state first that in both the Bardeen and ABG no-horizon spacetimes a static radius exist where an "antigravity" effect of the geometry is clearly demonstrated by vanishing of the angular momentum, $L_c = 0$, and the angular velocity, $\Omega_c = 0$, similarly to the case of the Kehagias-Sfetsos naked singularity spacetimes of the Ho\v{r}ava gravity \cite{Stu-Sche:2014:CLAQG:,Vie-etal:2014:PHYSR4:}, or the Reissner-Nordstrom naked sigularity spacetimes \cite{Stu-Hle:2002:ActaPhysSlov:,Pug-etal:2011:PHYSR4:}. At the static radius, an "antigravity" sphere occurs where test particles can remain in stable static equilibrium. No circular geodesics can exist under the stable static radius that is given by 
\begin{enumerate}
\item \emph{Bardeen} 
\beq
	r_{stat} = \sqrt{2} g
\eeq
\item \emph{ABG} 
\beq
	g^2 (g^2 - r^2) \sqrt{g^2 + r^2} +   (-2 g^4 - g^2 r^2 + r^4) = 0 . 
\eeq
\end{enumerate}
An exception is represented by the ABG no-horizon spacetimes with $g > 2$ where an inner Keplerian disc is placed under the stable static radius, having an outer edge at the unstable static radius corresponding to unstable equilibrium positions of static particles. Location of the static radii in dependence on the charge parameter of the Bardeen and ABG no-horizon spacetimes is demonstrated in Fig. 1. The antigravity sphere, being surrounded by a Keplerian disc, can be considered as a final state of the accretion process, and in some sense can represent an effective surface of the objects described by the Bardeen and ABG no-horizon spacetimes. 

Existence of the antigravity sphere is clearly related to the repulsive gravity acting very closely to the origin of coordinates. For $r \sim 0$, the lapse function of the Bardeen and ABG spacetimes can be written in a simple form 
\begin{enumerate}
\item \emph{Bardeen} 
\beq
	f(r \sim 0) \sim 1 - \frac{2}{g^3} r^2 ,
\eeq
\item \emph{ABG} 
\beq
	f(r \sim 0) \sim 1 - \frac{1}{g^2}(\frac{2}{g} - 1) r^2 , 
\eeq
\end{enumerate}
where we have to assume  $g>0$. Then we see that in the the Bardeen case the spacetime near the coordinate origin always demonstrates the gravitational repulsion of the same character as those related to the vacuum energy (the de Sitter type) \cite{Stu-Hle:1999:PHYSR4:}, where the cosmological term $\Lambda/3$ has to be replaced by $2/g^3$. In the ABG spacetimes, we obtain the gravitational repulsion of the de Sitter type( the vacuum energy) for $g < 2$ -- in this case, no circular geodesic orbits occur under the stable static radius where the antigravity sphere is located. On the other hand, for $g > 2$ the metric near the origin of coordinates is of the anti-de Sitter type with attrative gravity -- in this case circular geodesic orbits are possible even under the static radius. 

\subsection{Vanishing gradient of the angular frequency of circular geodesics}

The angular frequency of the circular geodesic motion, shortly, the Keplerian frequency, demonstrates a specific property in the spherically symmetric naked singularity spacetimes, both Reissner-Nordstrom, and Kehagias-Sfetsos, namely the change of sign of the gradient of the radial profile of the Keplerian frequency implying important consequences for the Keplerian accretion -- for details see \cite{Vie-etal:2014:PHYSR4:,Stu-Sche-Abd:2014:PHYSR4:,Stu-Sche:2014:CLAQG:}. It is quite interesting that this phenomenon occurs also in the case of the regular no-horizon spacetimes. 

The corresponding gradient of the Keplerian angular frequency is given by 
\begin{enumerate}
\item \emph{Bardeen}
	\beq
		\frac{\diff\Omega^2_c}{\diff r}=-\frac{3r(r^2-4g^2)}{(r^2+g^2)^{7/2}};
	\eeq
\item \emph{ABG}
	\beq
		\frac{\diff\Omega^2_c}{\diff r}=\frac{-3 r^5+4g^4r\left[3-2\sqrt{r^2+g^2}\right]+g^2r^3(9+4\sqrt{r^2+g^2})}{(r^2+g^2)^{9/2}}.
	\eeq
\end{enumerate}
The extrema of the function $\Omega_c^2(r;g)$ are located along the curves $r_{\Omega MAX}(g)$ determined by the condition $\diff\Omega^2_c/\diff r=0$ that for the particular spacetime imply the formulae
\begin{enumerate}
\item \emph{Bardeen}
	\beq
		r_{\Omega MAX/B} = 2g;
	\eeq
\item \emph{ABG} ($r_{\Omega MAX/ABG}$ determined in an implicit form)
	\beq
		-3  r^5 + 4 g^4 r (3  - 2 \sqrt{g^2 + r^2}) + 
		 g^2 r^3 (9  + 4 \sqrt{g^2 + r^2})=0.
	\eeq
\end{enumerate} 

The vanishing and change of the sign of the gradient of the Keplerian angular frequency radial profile has an important consequence for the Keplerian accretion discs, as the standard accretion governed by the magnetorotational  instability (MRI) requires decreasing of the Keplerian frequency with increasing radius, therefore, the radius $r_{\Omega MAX}$ can be considered as an inner edge of the standard Keplerian discs. Then an outer standard hot Keplerian disc can be continued by an inner cold Keplerian disc where the MRI cannot work \cite{Vie-etal:2014:PHYSR4:,Stu-Sche:2014:CLAQG:}. We postpone discussion of this phenomenon to a future work, as we focus our attention in the present paper on Keplerian rings radiating because of some processes being independent of the Keplerian accretion, as irradiation of the ring by an external sources. 

\subsection{Photon circular geodesics}

The photon circular geodesics are determined by the divergences of the $E_c$ and $L_c$ relations that can be put into the implicit form that reads 
\begin{enumerate}
\item \emph{Bardeen} 
\beq
	(r^2+g^2)^{5/2}-3 r^4=0, \label{phB}
\eeq
\item \emph{ABG} 
\beq
	(r^2+g^2)^{7/2}-r^4\left[-2g^2\sqrt{r^2+g^2}+3(r^2+g^2)\right]=0. \label{phABG}
\eeq
\end{enumerate}
The existence of the photon circular geodesics is allowed in the spacetimes with the charge parameter $g$ smaller than the critical charge parameter $g_{P}$ that is given in the following way
\begin{enumerate}
\item Bardeen 
	\beq
		g_{P/B} = 0.85865 ;
	\eeq
\item ABG 
	\beq
	    g_{P/ABG} = 0.690771 .
	\eeq
\end{enumerate} 
In the no-horizon spacetimes with the parameter $g < g_{P}$, two photon circular geodesics are allowed, the outer one being unstable relative to radial perturbations, and the inner one being stable. The results of the numerical calculations for the loci of the photon circular geodesics are presented in Fig. \ref{fig1}. Trapped photons move around the stable photon circular geodesic, just above the antigravity sphere at the stable static radius. 

\subsection{Marginally stable circular geodesics}

The inner edge of the Keplerian discs is located at the innermost stable circular orbit (ISCO) radius which is determined by simultaneously held conditions
\beq
	\frac{\diff V}{\diff r}=0\quad\textrm{and}\quad \frac{\diff^2 V}{\diff r^2}=0 .
\eeq 
Note that in the no-horizon spacetimes, similarly to the naked singularity spacetimes, also an outermost stable circular geodesic (OSCO) can exist, if the region of stable circular geodesics is separated by a region of unstable geodesics \cite{Vie-etal:2014:PHYSR4:,Stu-Sche:2014:CLAQG:}. Both ISCO and OSCO are thus marginally stable circular geodesics. The existence of unstable circular geodesics is allowed in the spacetimes with the charge parameter smaller that the critical charge $g_{S}$ that is given in the following way 
\begin{enumerate}
\item Bardeen 
	\beq
		g_{S/B} = 0.95629 ;
	\eeq
\item ABG 
	\beq
	    g_{S/ABG} = 0.74684 .
	\eeq
\end{enumerate} 
The results of the analysis of the stability of the circular geodesic motion and the loci of the marginally stable circular orbits are presented in Fig. 1. 

We can conclude that in the no-horizon regular Bardeen and ABG spacetimes with the spacetime parameter $g$ satisfying the conditions 
\begin{enumerate}
\item Bardeen 
	\beq
		g_{NoH/B} < g < g_{S/B} = 0.95629  ;
	\eeq
\item ABG 
	\beq
	    g_{NoH/ABG} < g < g_{S/ABG} = 0.74684 .
	\eeq
\end{enumerate}
two regions of stable circular geodesics exist, being separated by a region of unstable circular geodesics. The inner edge of the inner region of the stable circular geodesics is located at the static radius. In subregion of this parameter space region, given by 
\begin{enumerate}
\item Bardeen 
	\beq
		g_{NoH/B} < g < g_{P/B} = 0.85865  ,
	\eeq
\item ABG 
	\beq
	    g_{NoH/ABG} < g < g_{P/ABG} = 0.690771 ,
	\eeq
\end{enumerate}
two photon circular geodesics exist and no circular geodesics are allowed between the photon orbits. For the no-horizon spacetimes with the parameter $g$ satisfying the conditions  
\begin{enumerate}
\item Bardeen 
	\beq
		 g > g_{S/B} = 0.95629  ,
	\eeq
\item ABG 
	\beq
	     2 > g > g_{S/ABG} = 0.74684 ,
	\eeq
\end{enumerate}
only stable circular geodesics can exist, with the inner edge located at the stable static radius. 

In the case of the ABG no-horizon spacetimes with 
\beq
         g > 2 
\eeq
an exceptional situation arises, as an inner region of circular geodesics occurs under the stable static radius corresponding to the antigravity sphere -- this is not possible in the RN or Kehagias-Sfetsos naked singularity spacetimes. The inner geodesic region consists of stable orbits extending to the OSCO, above which unstable circular geodesics are located that terminate by the unstable static radius. 

Such a situation is clearly related to the behavior of the lapse function $f(r;g)$ at $r \sim 0$ when its form corresponds to the anti-de Sitter spacetime for $g > 2$ rather to the de Sitter spacetime that is relevant for $g < 2$. The attractive gravity at $r \sim 0$ enables existence of circular geodesics near the origin of coordinates in the spacetimes with $g > 2$, while the repulsive gravity forbids circular geodesics at $r \sim 0$ in the ABG no-horizon spacetimes with $g < 2$. The situation is clearly demonstrated by Figs \ref{fig1} and \ref{fig.3}. 

We have to stress that in the case of the no-horizon Bardeen spacetimes and the ABG spacetimes with $g < 2$, the situation is similar to those occuring in the spherically symmetric naked singularity spacetimes, as those described by the Reissner-Nordstrom geometry \cite{Stu-Hle:2002:ActaPhysSlov:,Pug-etal:2011:PHYSR4:} or the Kehagias-Sfetsos geometry \cite{Vie-etal:2014:PHYSR4:,Stu-Sche-Abd:2014:PHYSR4:,Stu-Sche:2014:CLAQG:}.

\subsection{Effective potential}

In order to clear up accretion phenomena related to the Keplerian accretion discs that are governed by the circular geodesic motion, the segments corresponding to stable circular geodesics, we give sequences of the effective potential taken for adequately chosen values of the specific angular momentum of test particles representing the accreting matter. \footnote{The Keplerian accretion onto general relativistic objects, namely Kerr black holes, has been first discussed in \cite{Nov-Tho:1973:BlaHol:,Pag-Tho:1974:ApJ:} However, the circular geodesics are relevant also for the toriodal accretion discs of perfect fluid governed by combined gravitational (inertial) forces and pressure gradients \cite{Koz-Abr-Jar:1978:ASTRA:,Stu-Sla-Hle:2000:ASTRA:}}. The stable circular geodesic orbits representing motion of matter accreting in a Keplerian disc from large distance (infinity) down to the central parts of the background spacetime are determined by the minima of the effective potential, and the edge of the Keplerian discs is determined by the ISCO. In the regular no-horizon spacetimes allowing for the existence of unstable circular geodesics, an inner Keplerian disc with the edge at OSCO, or at a lower radius, can be relevant too, similarly to the KS (and RN) naked singularity spacetimes \cite{Stu-Sche:2014:CLAQG:}. 

We can classify the Bardeen and ABG spacetimes according to the Keplerian accretion governed by the effective potential. Properties of the effective potential and the Keplerian accretion are governed by the geodesic structure of the spacetimes, namely by properties of the circular geodesics. 
\begin{figure}[ht]
\begin{center}
	\begin{tabular}{cc}
	\includegraphics[scale=0.7]{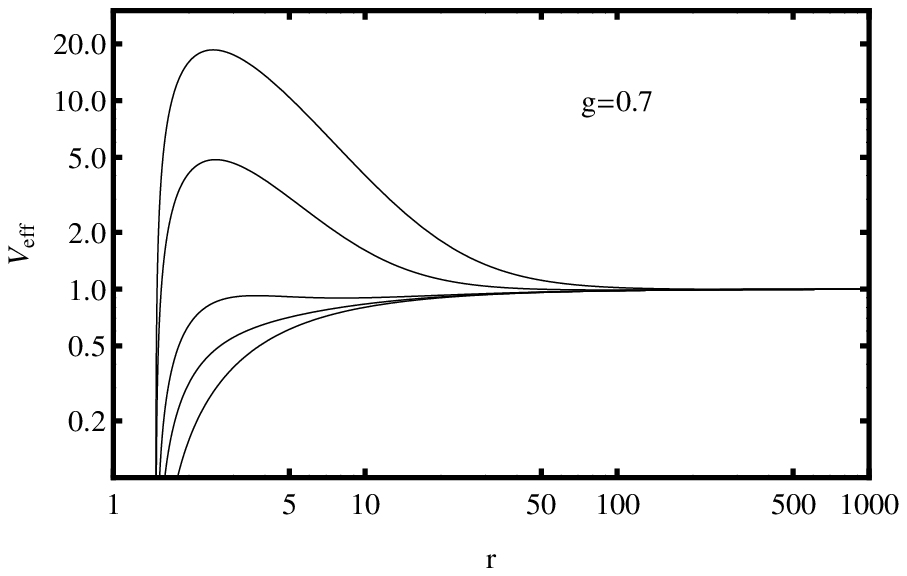}&\includegraphics[scale=0.7]{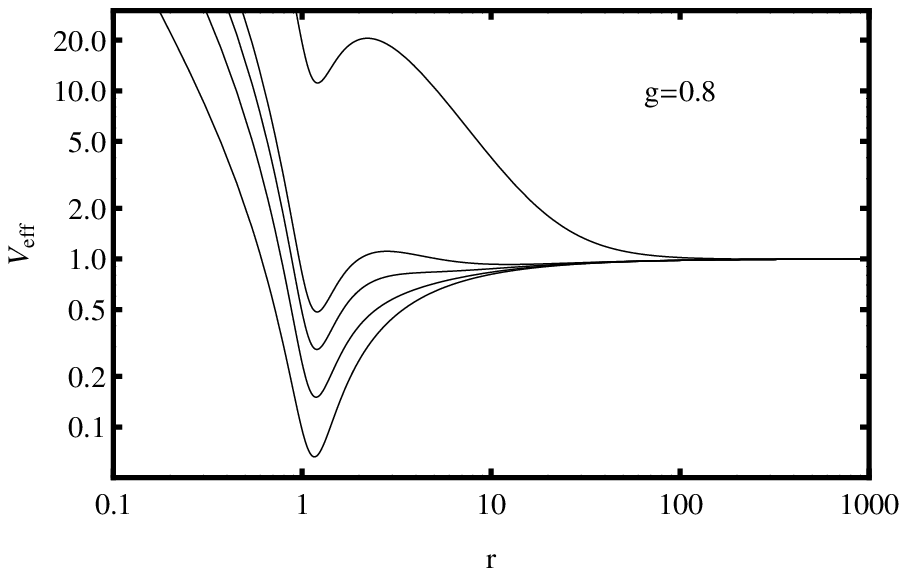}\\
	\includegraphics[scale=0.7]{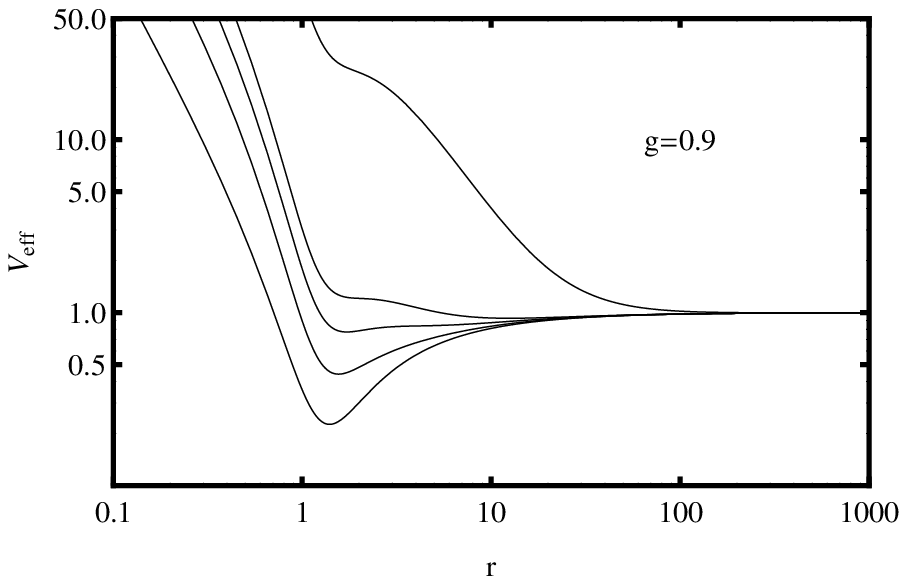}&\includegraphics[scale=0.7]{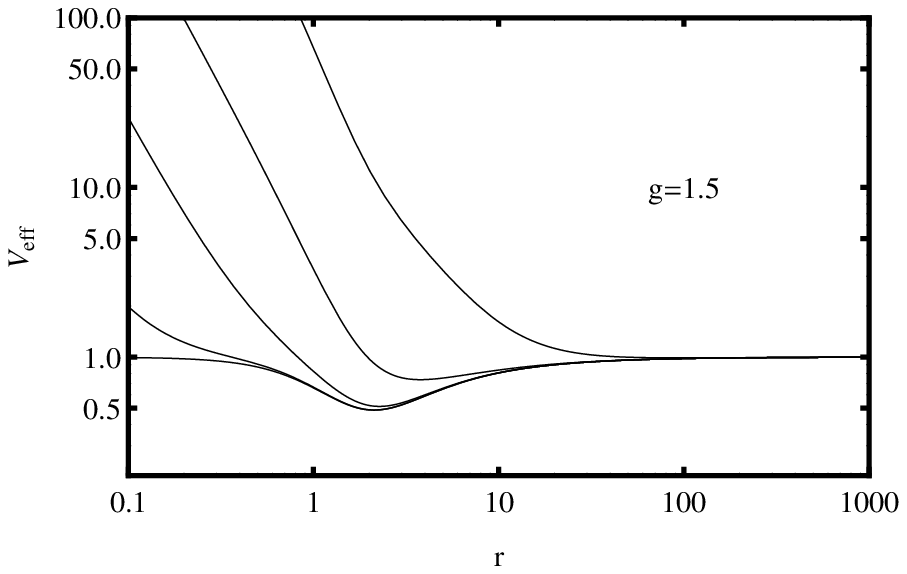}
	\end{tabular}
	\caption{Effective potential of massive test particles in the Bardeen spacetime is given for the four representative values of the charge parameter $g=0.7$, $0.8$, $0.9$, and $1.5$ (left to right, top to bottom). We assume $m=1$. The values of corresponding particle angular momentum are in the intervals  $L\in[0.1, 20.0]$ ($g=0.7$), $L\in[1.0, 20.0]$ ($g=0.8$ and $0.9$), $L\in[0.0, 10.0]$ ($g=1.5$). The Keplerian accretion is related to sequences of stable circular geodesics given by local minima of the effective potential. \label{fig.4}}
\end{center}
\end{figure}

\begin{figure}[ht]
\begin{center}
	\begin{tabular}{cc}
	\includegraphics[scale=0.7]{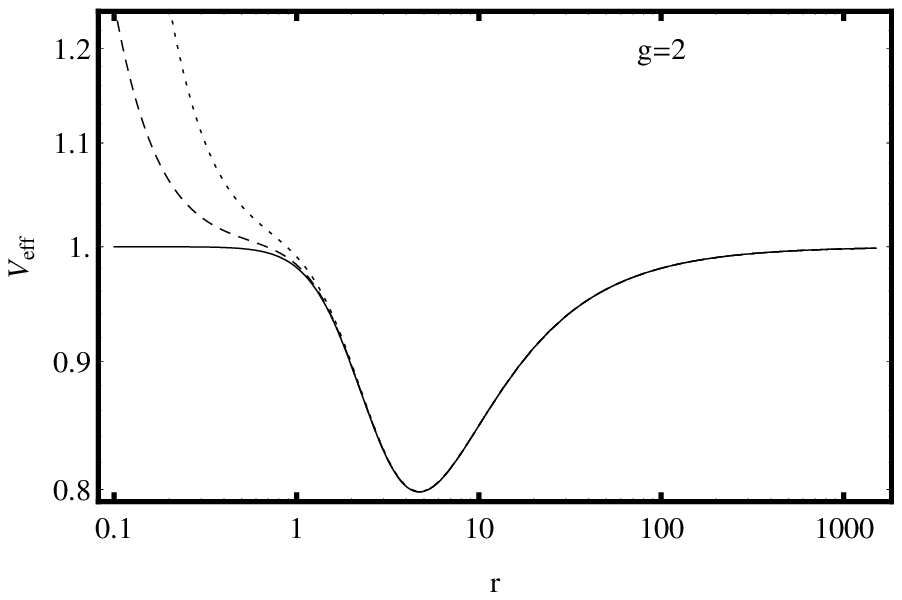}&\includegraphics[scale=0.7]{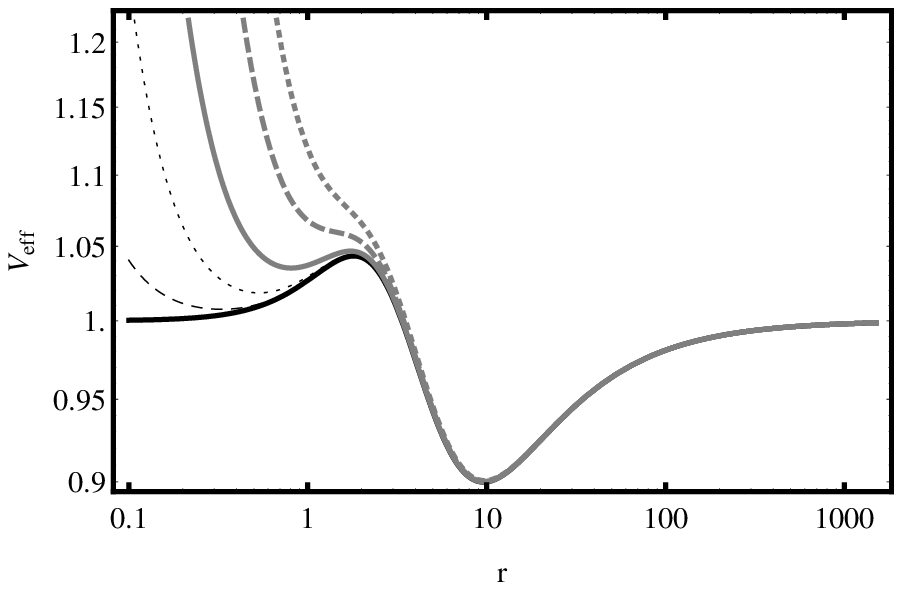}
	\end{tabular}
	\caption{Effective potential of massive test particles in the ABG spacetimes with two representative values of the charge parameter $g=2$(left) and $g=3$(right); we assume $m=1$. The values of corresponding particle angular momentum is $L=0.0$ (solid), $0.05$ (dashed), and $0.1$ (doted) in the case of $g=2$ and $L=0.$ (thick, black), $0.02$ (dashed), $0.05$ (doted), $0.1$ (thick,gray), $0.2$ (thick, dashed,gray), and $0.3$ (thick, dotted, gray) in the case of $g=3$. \label{fig.5}}
\end{center}
\end{figure}

The effective potential illustrates clearly the evolution of the angular momentum and energy of the accreting matter and demonstrates the possibility to obtain two Keplerian discs in the no-horizon spacetimes, similarly to the RN naked singularity spacetimes \cite{Stu-Hle:2002:ActaPhysSlov:,Pug-etal:2011:PHYSR4:}, or the KS naked singularity spacetimes of the modified Ho\v{r}ava gravity \cite{Vie-etal:2014:PHYSR4:,Stu-Sche-Abd:2014:PHYSR4:,Stu-Sche:2014:CLAQG:}. We give the characteristic $L=const$ sections of the effective potential governing the character of the Keplerian accretion for all the classes of the Bardeen and ABG black-hole and no-horizon spacetimes with qualitatively different behavior of the circular geodesics in Fig. \ref{fig.4}. 

The first one is related to the black-hole spacetimes, the three others are related to the no-horizon spacetimes -- spacetimes containing trapped photons and unstable circular geodesics, spacetimes admitting unstable circular geodesics, spacetimes admitting purely stable circular geodesics. For the ABG spacetimes with $g < 2$, the effective potential behaves in fully analogical way. 

For the ABG spacetimes with $g > 2$, behavior of the effective potential in the inner region of circular geodesics located under the stable static radius is illustrated in Fig. \ref{fig.5}. The inner Keplerian disc with stable circular orbits is located between the OSCO that is under the unstable static radius, and the inner edge at $r=0$. At the inner Keplerian disc both energy and angular momentum decrease with decreasing radius. On the other hand, the outer unstable part of the inner region of circular geodesics behaves unusually as both energy and angular momentum decrease with radius increasing up to the unstable static radius. 

The radial profiles of the specific angular momentum, and the specific energy of the Keplerian motion that are also helpful for understanding the Keplerian accretion are presented in Figs \ref{fig.2} and \ref{fig.3}. 

We do not discuss here the sign of the gradient of the angular velocity of the accreting matter, relevant in the accretion process governed by the MRI mechanism that requires $d\Omega_{c}^2/dr < 0$ \cite{Bal-Haw:1998:RevModPhys:}. This is crucial for the Keplerian discs and complicates strongly the situation of describing the accretion phenomena, as shown and discussed in \cite{Stu-Sche:2014:CLAQG:}. In the following, we concentrate attention to the Keplerian rings only in order to give local signatures of the strong gravity of the regular Bardeen and ABG spacetimes. In radiating Keplerian rings, the accretion process could be irrelevant. 

\section{Optical phenomena}

In this section the previous analysis is applied to show what fingerprints of each considered spacetime one can expect in the case of simple optical phenomena that, nevertheless, could be of observational relevance -- namely we shall consider the silhouette of the black hole and no-horizon spacetimes, and the spectral lines generated by orbiting Keplerian rings that are profiled by the strong gravity of the Bardeen and ABG spacetimes. 

\subsection{Escape cones and silhouette of regular black hole and no-horizon spacetimes}

The motion of photons in the equatorial plane of the spherically symmetric and static spacetimes is governed by the equation 
\begin{equation}
     \left[p^{r}\right]^{2}=E^{2}-f(r;g)\left(\frac{L^2}{r^2}\right).\label{eq_radial}
\end{equation}
The trajectories of photons are independent of energy, therefore, it is possible to determine the photon motion by their impact parameter 
\begin{equation}
                l = \frac{L}{E} ;
\end{equation}
then the turning points of the radial photon motion are given in terms of the effective potential related to the impact parameter by the relation 
\begin{equation}
             l^2 = V_{eff/ph}(r;g) \equiv \frac{r^2}{f(r;g)}. \label{eff/ph}
\end{equation}
We put for simplicity here and in the following $m=1$, i.e., the radius $r$, the parameter $g$, and the angular momentum (impact parameter) $L$ ($l$) are expressed in units of $m$. We illustrate the radial profile of the effective potential $V_{eff/ph}(r;g)$ for representative choices of the parameter $g$ in Fig. 6. Clearly, in the no-horizon spacetimes allowing for existence of circular photon geodesics, the trapped photons exist only if radiated by sources located under the outer unstable photon circular orbit. All photons radiated above the radius of the outer photon circular orbit escape to infinity except those winding up at the photon circular orbit, having $l=l_{ph(u)}$. In the no-horizon spacetimes having no photon circular orbits, all photons escape to infinity if radiated from any position in the spacetime, except those moving inwards purely radially that are terminating at $r=0$. 
\begin{figure}
\begin{center}
\begin{tabular}{cc}
\includegraphics[scale=0.7]{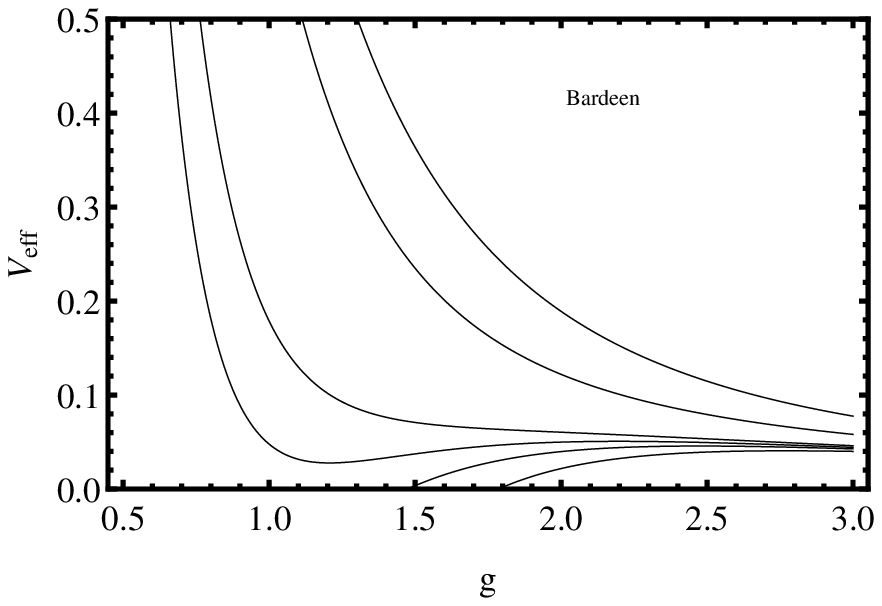}&\includegraphics[scale=0.7]{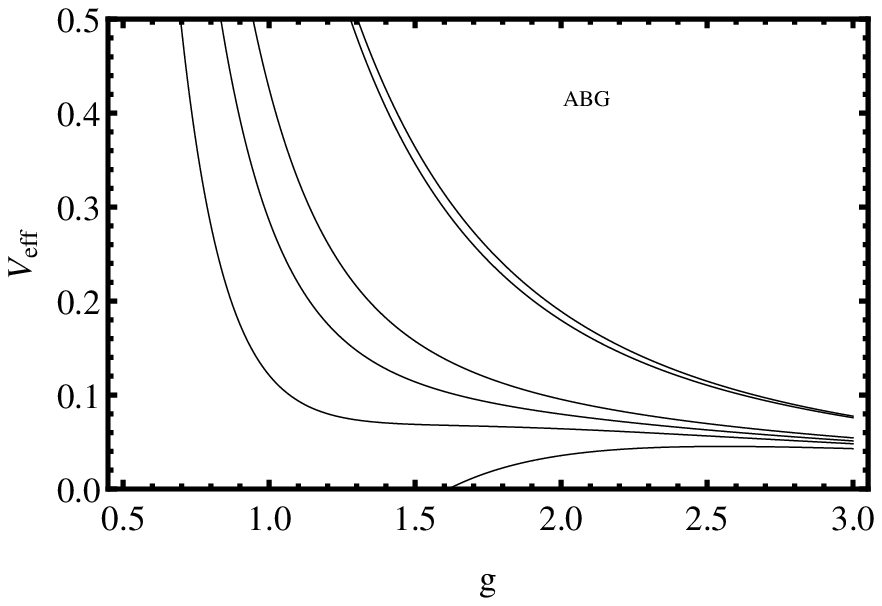}
\end{tabular}
\caption{Effective potential of the photon motion is given in the Bardeen (left) and ABG (right) spacetimes for five representative values of the charge parameter $g=0.5$, $0.7$, $0.8$, $0.9$, $1.5$, and $2.5$ (bottom curves to the upper one).\label{fig.6}  }
\end{center}
\end{figure}

The photon circular orbits, i.e., their radius $r_{ph}(g)$ and the related impact parameters $l_{ph}(g)$ can be found from the condition
\begin{equation}
\frac{dV_{eff/ph}}{dr}=0 . \label{eq_radil_null}
\end{equation}
The photon circular orbits are located at radii $r_{ph}$ satisfying Eq. (\ref{phB}) for the Bardeen spacetimes, and Eq. (\ref{phABG}) for the ABG spacetimes. For the Bardeen (ABG) black-hole spacetimes with $g < g_{NoH/B}=0.7698$ ($g < g_{NoH/ABG}=0.6342$), one unstable photon circular orbit exists, while for the Bardeen (ABG) no-horizon spacetimes with $g_{NoH/B}=0.7698 < g < g_{P/B}=0.85865$ ($g_{NoH/ABG}=0.6342 < g < g_{P/ABG}=0.69077$) two photon circular geodesics exist, the inner one being stable relative to radial perturbations, and the outer one being unstable. 

\begin{figure}
\begin{center}
\includegraphics[scale=0.9]{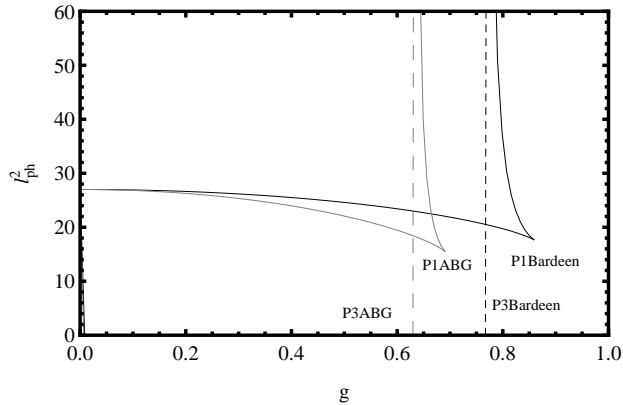}
\caption{The impact parameter corresponding to the photon circular orbits given as function of parameter $g$. The lower branch corresponds to the unstable (outer) photon circular geodesics, the upper branch corresponds to the stable (inner) photon circular geodesics.\label{fig.7}}
\end{center}
\end{figure}

The corresponding value of the impact parameter $l_{ph}$ of the circular photon orbits follows from Eq. (\ref{eff/ph}), i.e., there is 
\begin{equation}
l_{ph}^{2}=\frac{r_{ph}^{2}}{f(r_{ph};g)}, 
\end{equation}
where $r_{ph} = r_{ph-s}$ or $r_{ph} = r_{ph-u}$ for the inner and outer photon orbits in the no-horizon spacetimes. The values of the impact parameter of the photon circular orbits are given as a function of the charge parameter $g$ in Fig. 7. We present the dependence of the impact parameter for both the stable and unstable circular photon orbits. However, for distant observers only the unstable orbit is the relevant one. 

For the extreme Bardeen (ABG) black holes with $g = g_{NoH/B}=0.7698$ ($g = g_{NoH/ABG}=0.6342$), the photon circular orbit is located at the radius $r_{ph(extr)/B}=2.301$ ($r_{ph(extr)/ABG}=2.136$), and the related impact parameter takes the value 
\begin{equation}
                      l_{ph(extr)/B}\doteq 4.524, 
\end{equation}
\begin{equation}                      
                      l_{ph(extr)/ABG}\doteq 4.275.
\end{equation}
In the limiting Bardeen (ABG) no-horizon spacetimes allowing for existence of the photon circular geodesics, with $g = g_{P/B}=0.85865$ ($g = g_{P/ABG}=0.69077$), the photon circular orbit is located at $r_{ph(cr)/B}=1.72$ ($r_{ph(cr)/ABG}=1.57$), and the value of the corresponding impact parameter reads 
\begin{equation}
                      l_{ph(cr)/B}\doteq 4.206, 
\end{equation}
\begin{equation}                     
                      l_{ph(cr)/ABG}\doteq 3.937.
\end{equation}
Recall that in the vacuum Schwarzschild spacetime, the photon circular orbit is located at $r_{ph}=3$, and the impact parameter $l_{ph}=3\sqrt{3}=5.196$. 

The opening angle of the escaping cones is governed by the unstable photon circular geodesics. If the opening angle is related to the static reference frames is the spherically symmetric static spacetimes, it is determined by the simple formula \cite{Stu:1983:BAC:,Stu-Cal:1991:GenRelGrav:,Stu-Hle:1999:PHYSR4:}
\begin{equation}
    \alpha_{esc} =\arcsin\left[ \frac{\sqrt{f(r;g)}}{r}l_{ph(u)}\right]
\end{equation}
where $l_{ph-u}(g)$ is the impact parameter of the unstable photon circular orbit. The silhouette of a black hole or a no-horizon spacetime is determined by the opening angle of the escaping cone related to a distant static observer \cite{Stu-Hle:1999:PHYSR4:}, as the opening angle determines the photons trapped by the black hole or the no-horizon spacetime. In the no-horizon spacetimes, only the photons with $l=l_{ph-u}$ are trapped by the gravitational field and form a circular silhouette, while in the black hole spacetimes, the photons with impact parameter $l<l_{ph}$ are captured and give the silhouette of the black hole filling interior of the circle. No silhouette ocurrs for the no-horizon spacetimes not allowing for existence of photon circular orbits. The distant-observer sky represented by the standard parameters $[\alpha,\beta]$ \cite{Bar:1973:BlaHol:,Sche-Stu:2008:IJMPD:}, related to the impact parameters of the outcoming photons, then reflects a dark circle of the radius governed by the value of the impact parameter $l_{ph-u}$ of photons trapped on the photon orbit that determines the silhouette. Extension of the circular silhouette is determined by the impact parameter $l_{ph-u}$ -- see Fig.7. Clearly, the silhouette of the black hole regular spacetimes has to be smaller than the silhouette of the Schwarzschild black holes. The circles giving the silhouette of the no-horizon spacetimes have to be smaller than those of the regular black holes. For a given dimensionless charge parameter $g/m$, extension of the silhouette of the Bardeen spacetime has to be always larger than those of the ABG spacetime. In fact, it could be possible to distinguish the Bardeen and ABG spacetimes, if the precise measurements planned for observations of the Sgr A* Galaxy source will be realized under proper conditions enabling observability of the central parts of the spacetimes. 

Of course, in realistic situations, when an antigravity sphere will be created because of accretion processes, we have to assume that the sphere is absorbing all the incoming radiation. Then the silhouette of the regular, no-horizon spacetimes of the Bardeen and ABG type will be given by the antigravity sphere with extension $R \sim r_{stat}$ rather than by the unstable photon orbits of the spacetimes. On the other hand, the antigravity sphere could be itself a source of significant radiation. 

\subsection{Frequency shift}

In the equatorial plane, each photon emitted by the Keplerian ring suffers from combined gravitational and Doppler frequency shift which takes the general form
\beq
	1+z=\frac{\sqrt{f-r^2\Omega_{c}^2}}{1-l \Omega_{c}},
\eeq 
where the photon impact parameter $l=-p_\phi/p_t=L_z/E$ and $\Omega_{c}$ is the Keplerian angular frequency. 

In the special case of radiation from the stationary sources at the static radius (antigravity sphere) the frequency shift takes the simple form 
\beq
	1+z=\frac{\nu_{o}}{\nu_{e}} = \sqrt{-g_{tt}(r_{stat};g)} = \sqrt{f(r_{stat},g)} ,
\eeq 
We show the dependence of the redshift of radiation emitted by static sources at the static radius in dependence on the charge parameter $g$ in Fig.8 for both the Bardeen and ABG spacetimes. In the case of the ABG spacetimes with $g > 2$ we give the frequency shift also for the inner static radius where unstable equilibrium of test particles is possible. We can see that for the stable equilibrium positions there is always a redshift of the photons. For a given charge parameter $g$, the redshift is substantially stronger for the Bardeen spacetimes in comparison to the ABG spacetimes. This could be clearly an interesting observational signature of these spacetimes. Notice that photons from the unstable static radius of the ABG spacetimes with $g > 2$ are blueshifted. 
\begin{figure}
\begin{center}
\includegraphics[scale=0.8]{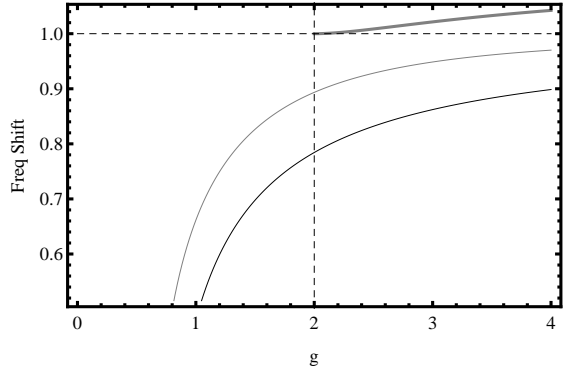}
\caption{The frequency shift of photons emitted by tatic observers at the static radii is illustrated as a function of the charge parameter $g$ for the Bardeen spacetimes (black solid line) and the ABG spacetimes (gray lines).   \label{fig.8}}
\end{center}
\end{figure}

\subsection{Radiative flux}

The specific flux at a detector, $F_{\nu0}$, is constructed by binning the photons (pixels) contributing to the specific flux $F$ at the observed frequency $\nu_0$. Let i-th pixel on the detector subtends the solid angle $\Delta\Pi_i$, then the corresponding flux $\Delta F_i(\nu_0)$ reads 
\beq
	\Delta F_i(\nu_o)=I_{o}(\nu_o)\Delta\Omega_i=g_i^3 I_e(\nu_o/g_i) \Delta\Pi_i,
\eeq
where the specific intensity of naturally (thermally) broadened line with the power law emissivity model is given by 
\beq
	I_e=\epsilon_o r^{-p} \exp[-\gamma(\nu_o/g -\nu_0)^2].
\eeq
In our simulations the dimensionless parameter $\gamma=10^3$. 
The solid angle is given by the coordinates $\alpha$ and $\beta$ on the observer plane due to the relation $d\Pi =  d\alpha d\beta/{D^{2}_{\rm o}}$, where $D_{\rm o}$ denotes the distance to the source -- for details see \cite{Bar:1973:BlaHol:,Sche-Stu:2008:IJMPD:}. The coordinates $\alpha$ and $\beta$ can be then expressed in terms of the radius $r_{\rm e}$ of the source orbit and the related redshift factor $1+z=\nu_{\rm o}/\nu_{\rm e}$. The Jacobian of the transformation $(\alpha,\beta) \rightarrow (r_{\rm e},g)$ implies \cite{Sche-Stu:2008:GenRelGrav:,Sche-Stu:2013:JCAP:}
\begin{equation}
d\Pi=\frac{q}{D_{\rm o}^2\sin\theta_{\rm o}\sqrt{q-\lambda^2\cot^2\theta_{\rm o}}}\left|\frac{\partial r_{\rm e}}{\partial\lambda}\frac{\partial g}{\partial q}-\frac{\partial r_{\rm e}}{\partial q}\frac{\partial g}{\partial \lambda}\right|^{-1}\!\!\!\!\!{d} g{d} r_{\rm e} \rightarrow \Delta\Pi_i.
\end{equation}
The parameter $q$ represents the total photon impact parameter related to the plane of motion of the photon, while $\lambda$ represents the axial impact parameter related to the plane of the Keplerian ring.
 
To obtain the specific flux at a particular frequency $\nu_o$, all contributions given by $\Delta F_i(\nu_o)$ are summed to obtain 
\beq
	F(\nu_o)=\sum_{i} \Delta F(\nu_o)_i. 
\eeq  
Details of the construction of the profiled spectral lines in the black hole spacetimes can be found in \cite{Sche-Stu:2008:GenRelGrav:,Sche-Stu:2008:IJMPD:}, and in the case of the spacetimes having no horizons, it is treated in \cite{Stu-Sche:2010:CLAQG:,Stu-Sche:2012:CLAQG:,Stu-Sche:2013:CLAQG:,Stu-Sche-Abd:2014:PHYSR4:,Stu-Sche:2014:CLAQG:}. 

\subsection{Profiles of spectral lines radiated from Keplerian rings}

In order to demonstrate effect of the Bardeen and ABG spacetimes on radiation governing the observational phenomena, we construct series of profiled spectral lines originating from Keplerian rings located at limiting radii characteristic for the geodesic circular motion, namely the marginally stable orbits, and the radius of vanishing gradient of the Keplerian angular velocity at the spacetimes allowing only for the stable circular orbits. The spectral lines are constructed for three characteristic values of the inclination of the radiating ring to distant observer, $\theta_o = 30^o, 60^o, 85^o$. For comparison, the profiled line created in the Schwarzschild spacetime at the ISCO is presented in all cases. 

\subsubsection{Bardeen spacetimes}
In Fig.9 we present the profiled spectral lines generated by the Keplerian rings located at $r_{ISCO}$ of the outer region of circular geodesics in the spacetimes with $g < g_{S/B}$, and at the $r_{\Omega MAX}$ in the spacetimes with $g > g_{S/B}$. We give the profiled lines for a black hole  spacetime, and all three classes of no-horizon spacetimes, namely those allowing for the stable and unstable photon geodesics, those allowing for both stable and unstable circular geodesics, but no photon circular orbits, and for those allowing only stable circular geodesics. 

We can see that the characteristic double-horned profile caused by the Doppler shift with blue horn larger than the red horn is enriched by additional small horns occuring for the mediate ($60^\circ$) and large ($85^\circ$) inclination angles in the no-horizon spacetimes admitting existence of unstable circular geodesics. In the case of small and
mediate inclination angles, the profiled lines of the Bardeen spacetimes are redshifted relative to the profiled line generated in the Schwarzschild spacetime, at both red and blue end of the Schwarzschild line. The frequency shift is smallest for the black hole case, and in the no-horizon spacetimes the shift increases with increasing parameter $g$. Extension of the profiled line is smallest for the spacetime admitting only the stable circular orbits. For large inclination angle, the extension of the profiled line is again smallest for the no-horizon spacetimes admitting only the stable circular geodesics when it is given for the radius of vanishing Keplerian angular frequency gradient. However, extension of the profiled line increases with increasing parameter $g$ in the spacetimes admitting for existence of ISCO, being smallest for the black hole spacetimes. For large inclination angle ($85^\circ$), extension of the Bardeen profiled line is larger than those generated in the Schwarzschild spacetime in both red and blue edge of the frequency range. The shift is quite strong at the blue end where it reaches value of $1+z \sim 1.5 - 1.6$, in comparison to the Schwarschild value of $1+z \sim 1.4$, giving thus a clear observational imprint of the no-horizon Bardeen spacetimes, along with the additional humps in the profile of the spectral line. 

In Fig.10 we compare for completeness the profiled spectral lines created at the OSCO of the inner region of circular geodesics in the Bardeen spacetimes with $g_{P/B} < g < g_{S/B}$ admitting existence of unstable circular geodesics, with those created at the ISCO of the outer region of circular geodesics. Now we observe an extremely strong dependence on the inclination angle, when strong additional humps occur in the profile, having a doubled-horn like character for mediate and large inclination angles. For small inclination angle ($30^\circ$), whole the profiled line generated at the OSCO is shifted to the red end of the spectrum in comparison with the profiled line generated at the ISCO. For large inclination angle ($85^\circ$), the OSCO line is much more extended in comparison to the ISCO line at both blue and red end, keeping the doubled-horn character. At the blue end the frequency shift goes to extremely large value of  $1+z \sim 2.5$, giving thus extremely profound signature of the presence of a radiating inner Keplerian ring. 

\subsubsection{ABG spacetimes}
In Fig.11 we present the profiled lines generated by the Keplerian rings located at the $r_{ISCO}$ of the outer region of circular geodesics in the ABG spacetimes with $g < g_{S/ABG}$, and at the $r_{\Omega MAX}$ in the ABG spacetimes with $g > g_{S/ABG}$. We give the profiled spectral lines for a black hole spacetime, and all three classes of the no-horizon spacetimes, as in the Bardeen spacetimes. Now the role of the inclination angle is little bit different in comparison to the Bardeen spacetimes. The difference is most profound for the spacetimes admitting only the stable orbits when the Keplerian ring is assumed at $r_{\Omega MAX}$ -- for all inclination angles the profiled line is located inside the range of frequencies corresponding the the Schwarzschild ISCO line, and for the large inclination angle ($\theta_0 =85^\circ$) even doubled-horn structure occurs with the central horns exceeding the horns at the red and blue edge of the profiled line. This is an exceptional case that does not occur in no other configuration of the Keplerian rings located in the ABG or Bardeen regular spacetimes. The profiled lines generated at the ISCO position of the Keplerian ring demonstrate quantitative differences in comparison to the situation in the Bardeen spacetimes. Generally, their range is more strongly redshifted relative to the Schwarzschild ISCO line, and the ABG black hole line is rather close to the Schwarzschild line for all inclination angles. In the no-horizon spacetimes, the profiled line is always located under the blue edge of the Schwarzschild line for all inclination angles, contrary to the case of the Bardeen spacetimes. Only the ABG black hole line slightly exceeds the Schwarzschild line at the blue edge. It is thus clear that the profiled lines generated by radiating Keplerian rings at some significant positions at the ABG spacetimes can be well distinguished from those generated at the Bardeen spacetimes, if the inclination angle is known. 

The profiled spectral lines created in the ABG spacetimes with $g_{P/ABG} < g < g_{S/ABG}$ at the OSCO of the inner region of circular geodesics are compared with those created at the ISCO of the outer region of circular geodesics in Fig.12. In this case, no doubled-horn spectral line structure is obtained in the case of all the inclination angles under consideration, demonstrating thus again a clear difference in comparison to the situation in the Bardeen spacetime reflected in Fig.10. For all the inclination angles the line generated at the OSCO is strongly shifted to the red edge of the spectrum and is significantly flatter in comparison to the line generated at the ISCO, giving thus another strong difference of the Bardeen and ABG spacetimes. 

In the special case of the ABG spacetimes with $g > 2$, we illustrate for completeness also the properties of the inner region of circular geodesics located under the stable static radius. We construct the profiled lines radiated by Keplerian rings at the OSCO, at the $r_{\Omega MAX}$, and for completeness also at the $1.5\times r_{\Omega MAX}$. We demonstrate that the shape of the profiled line is the same for all the inclination angles and the ring located at OSCO, but its position relative to the Schwarzschild ISCO line changes significantly with the inclination angle. For small angle ($30^o$), the OSCO line extends outside the frequency range of the Schwarzschild ISCO line, being shifted to the blue end. For mediate and large inclination angles, the profiled line extends inside the frequency range of the Schwarzschild line, with extension slightly increasing with inclination angle increasing. Similar situation occurs also for the Keplerian rings located at $r_{\Omega MAX}$ and at $1.5 \times r_{\Omega MAX}$, but in these cases an additional humpy structure appear at the central region ($1+z \sim 1$) of the profiled lines observed under the large inclination angle. Of course, we have to stress that probability to observe the profiled lines from the region hidden under the stable static radius where a stable antigravity sphere
can be expected is very low. 

It has to be said that both the shape and extension of the profiled spectral lines generated by Keplerian rings in the regular Bardeen or ABG spacetimes differs from those generated in the Kehagias-Sfetsos spacetimes that were constructed in \cite{Vie-etal:2014:PHYSR4:}, or those occuring in the Kerr naked singularity spacetimes \cite{Sche-Stu:2013:JCAP:}. We can conclude that the regular Bardeen or ABG spacetimes could give observational signatures that can be clearly distinguished from those generated in vicinity of other exotic sources, as the Kerr or Kehagias-Sfetsos naked singularities. 

\begin{figure}[H]
\begin{center}
\begin{tabular}{cc}
\includegraphics[scale=0.7]{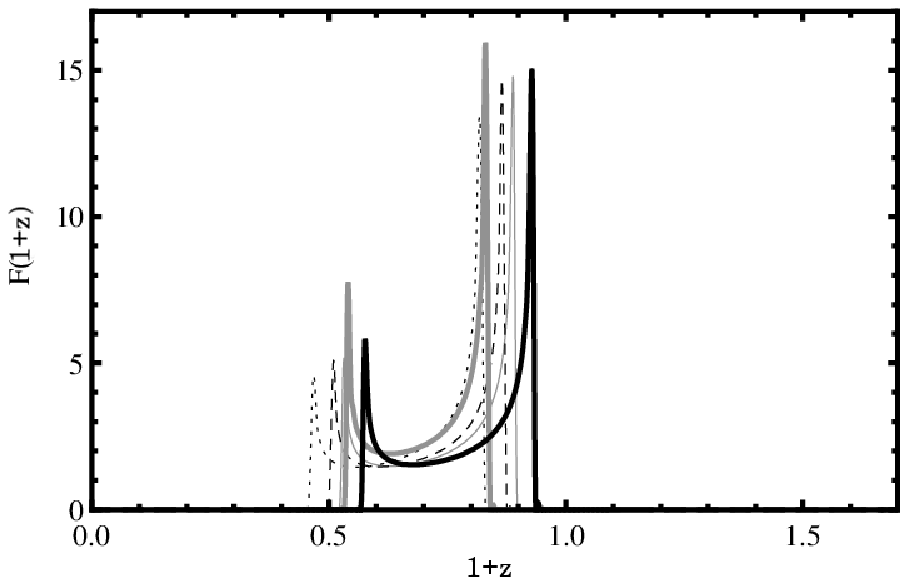}&\includegraphics[scale=0.7]{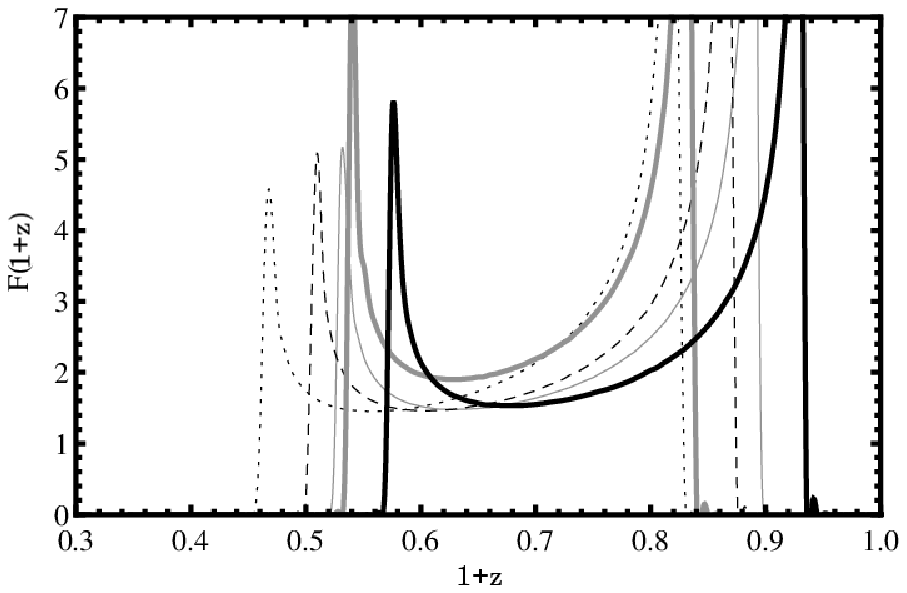}\\
\includegraphics[scale=0.7]{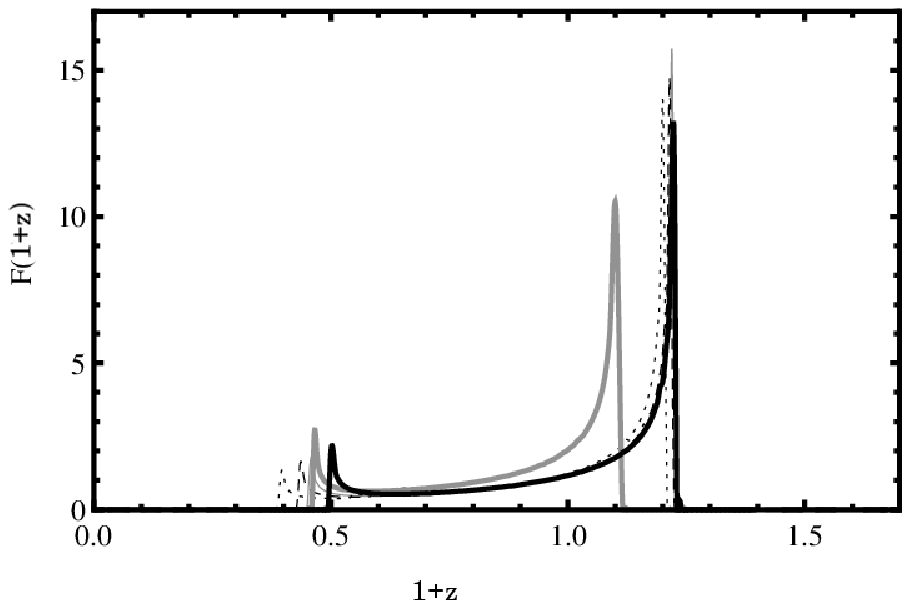}&\includegraphics[scale=0.7]{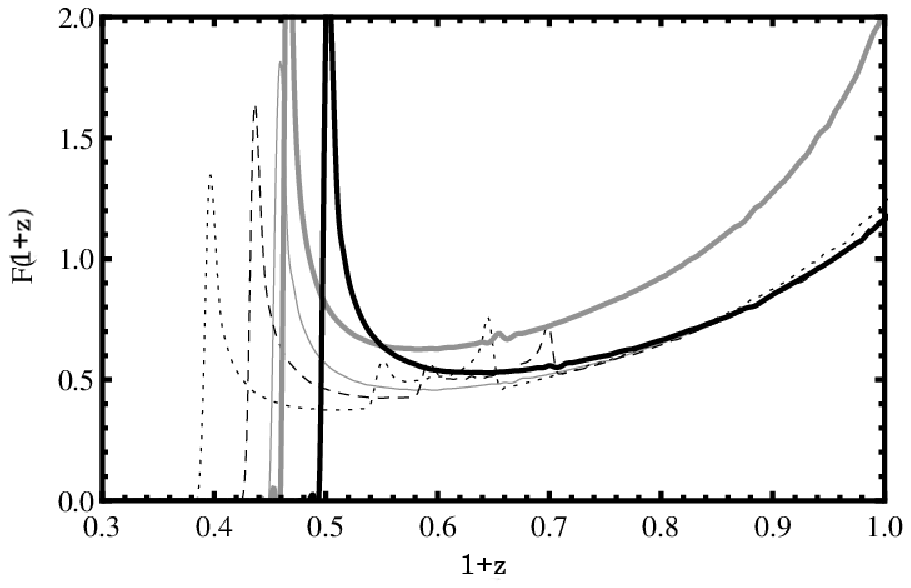}\\
\includegraphics[scale=0.7]{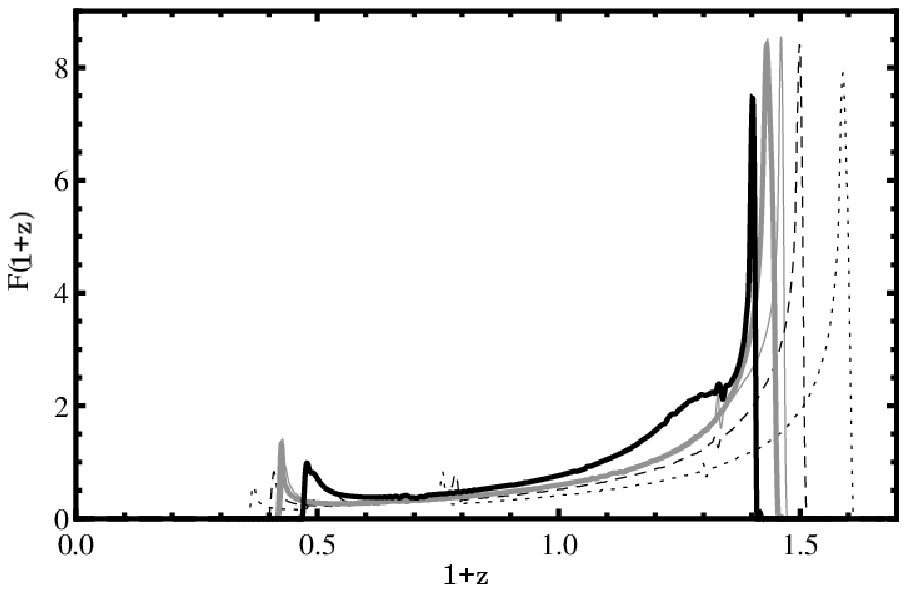}&\includegraphics[scale=0.7]{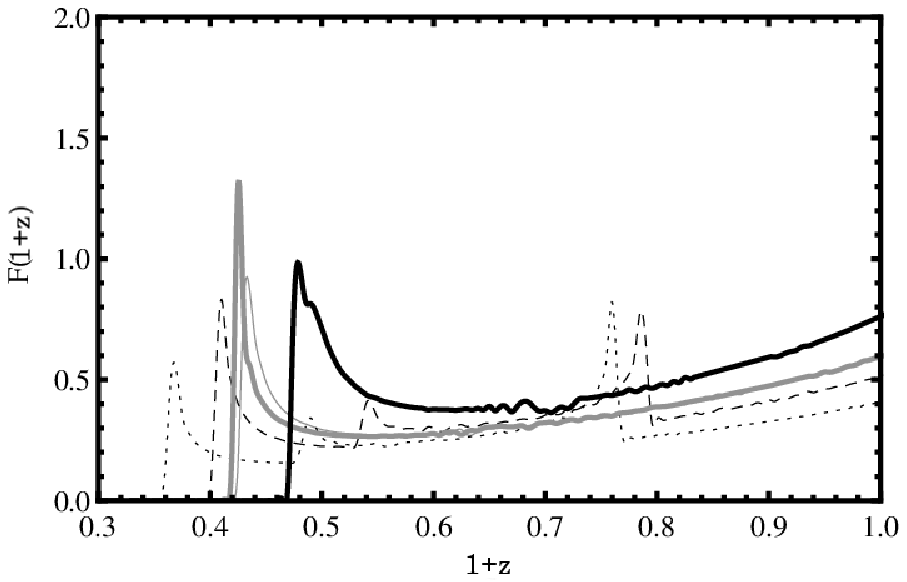}
\end{tabular}
\caption{Profiles of the spectral lines generated by Keplerian rings in the \emph{Bardeen} spacetimes with the charge parameter $g = 0.7$ (solid,gray), $0.8$ (dashed,black), $0.9$ (dotted,black), and $1.5$ (thick,gray) are compared to those generated in the Schwarzschild spacetime (thick, black). The profiles were constructed for the
three representative values of observer inclination $\theta_o = 30^\circ$ (top), $\theta_o = 60^\circ$
(middle), and $\theta_o = 85^\circ$ (bottom). The location of the Keplerian ring in the \emph{Bardeen} spacetimes is at $r_{ISCO}$ (up to $g = 0.9$), $r_{\Omega MAX}$ (for $g=1.5$), and in the Schwarzschild spacetime it is at $r_{ISCO} = 6$. In the right column, magnification of the red end of the profiled lines is presented in order to give details of their small humps.\label{fig.9}}
\end{center}
\end{figure}

\begin{figure}[H]
\begin{center}
\begin{tabular}{cc}
\includegraphics[scale=0.7]{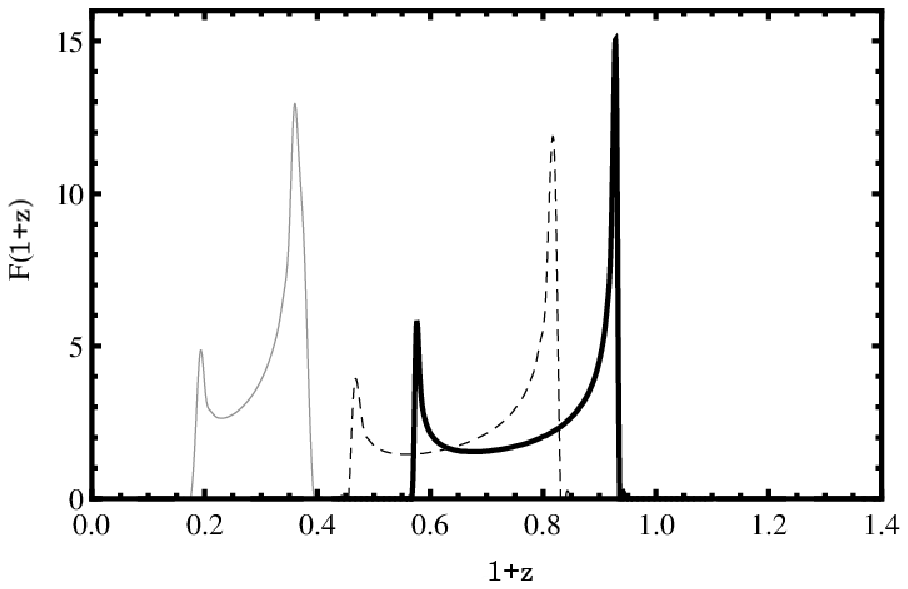}&\includegraphics[scale=0.7]{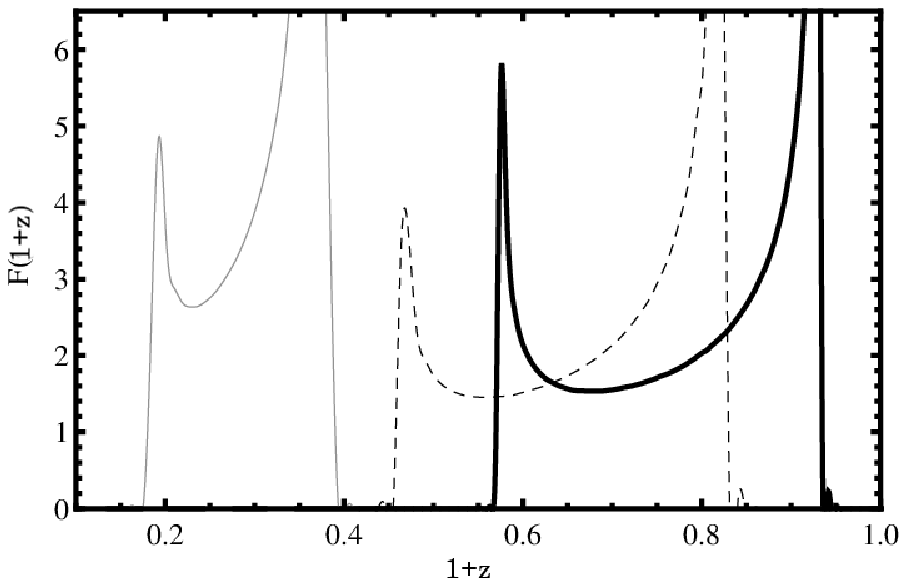}\\
\includegraphics[scale=0.7]{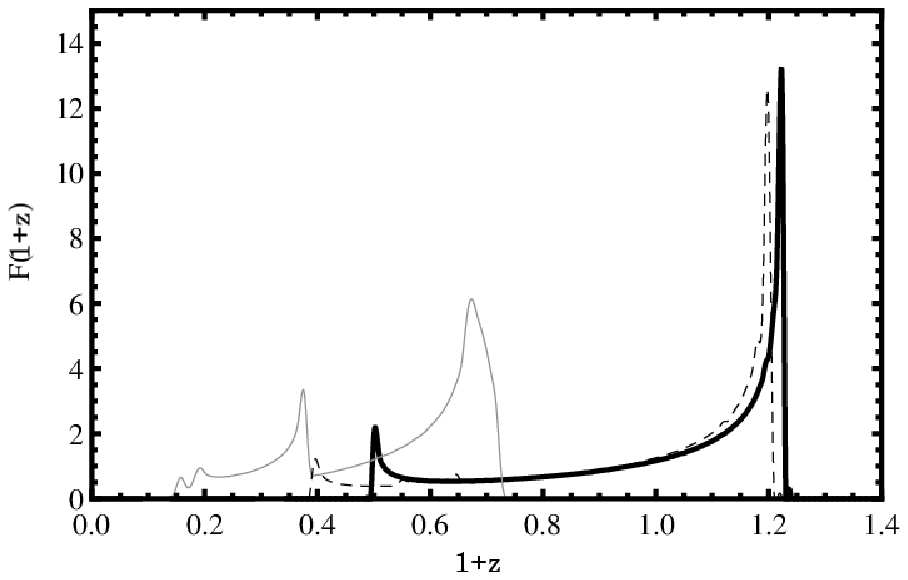}&\includegraphics[scale=0.7]{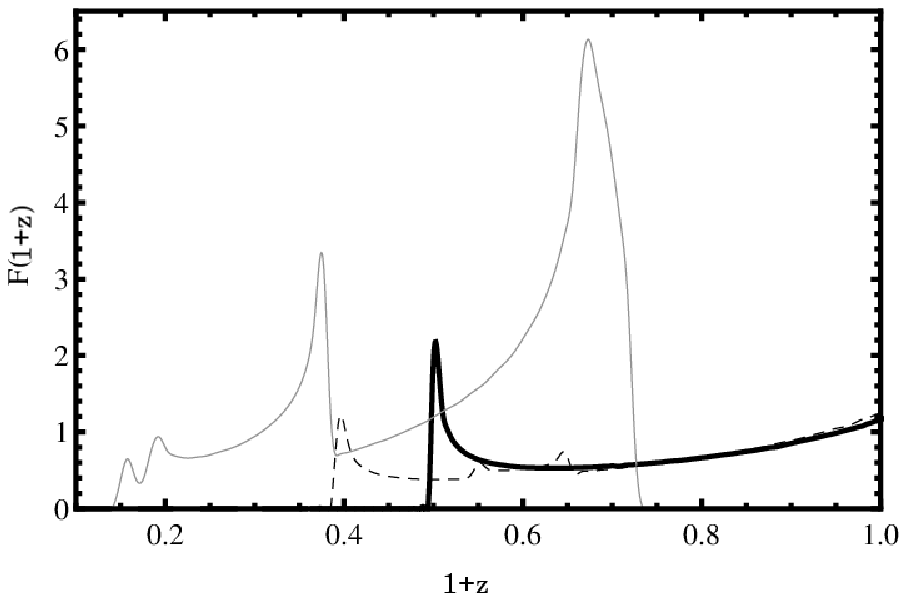}\\
\includegraphics[scale=0.7]{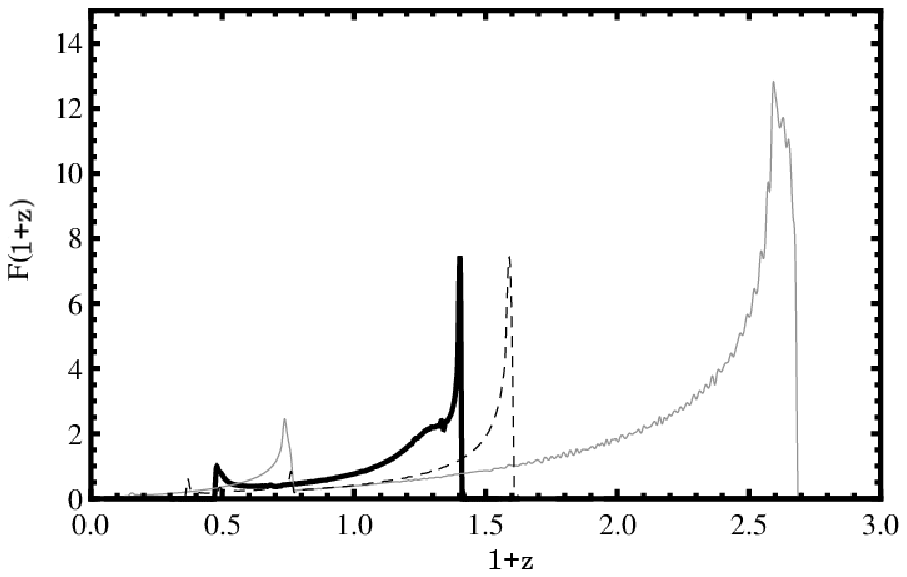}&\includegraphics[scale=0.7]{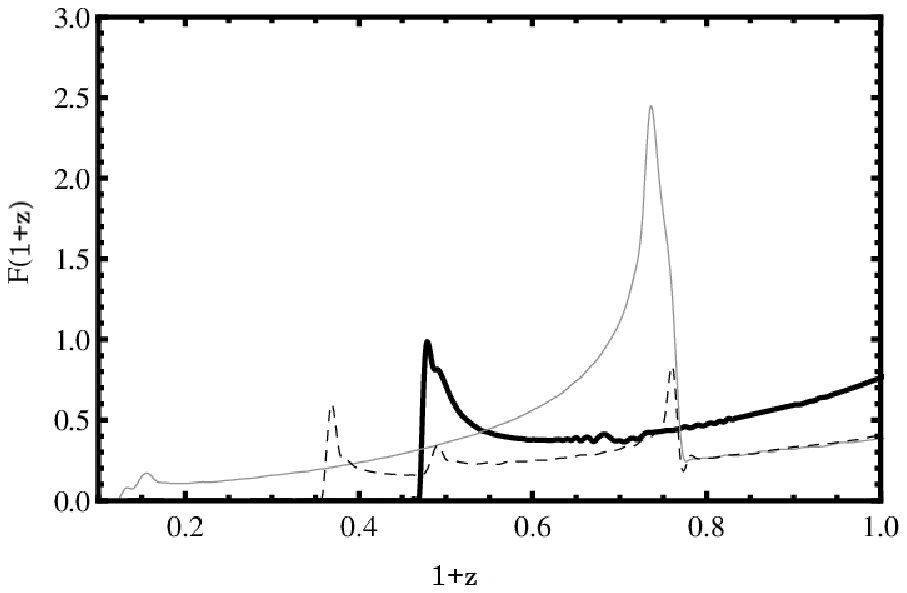}
\end{tabular}
\caption{Profiles of the spectral lines generated by Keplerian rings located in the \emph{Bardeen} spacetime with the charge parameter $g=0.9$ at the OSCO of the inner region of the stable circular geodesics (gray) and at the ISCO of the outer region of stable circular geodesics (dashed). They are compared to the line generated at the ISCO in the Schwarzschild spacetime (thick, black). The profiles were constructed for the three representative values of observer inclination $\theta_o=30^\circ$(top), $\theta_o=60^\circ$ (middle), and $\theta_o=85^\circ$ (bottom). \label{fig.10} }
\end{center}
\end{figure}

\begin{figure}[H]
\begin{center}
\begin{tabular}{cc}
\includegraphics[scale=0.7]{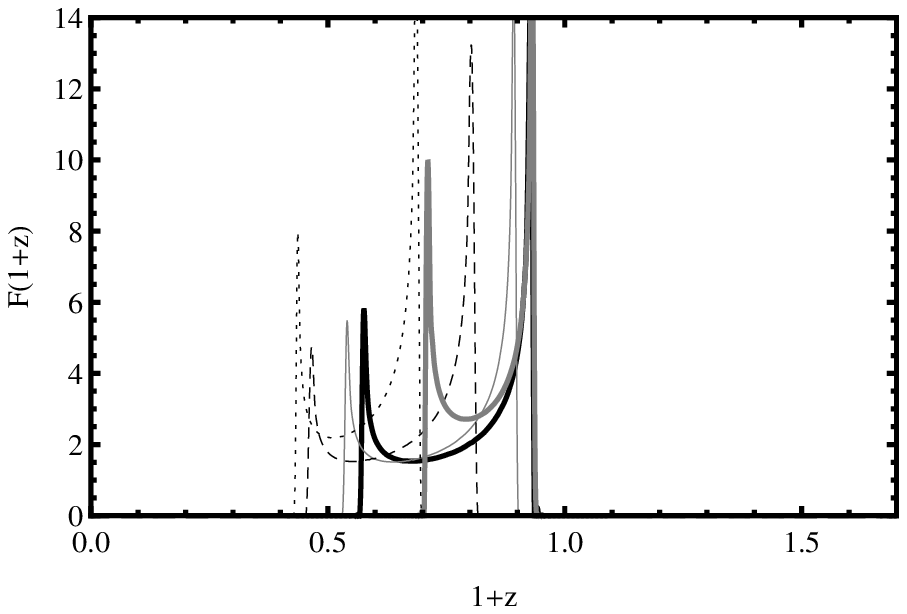}&\includegraphics[scale=0.7]{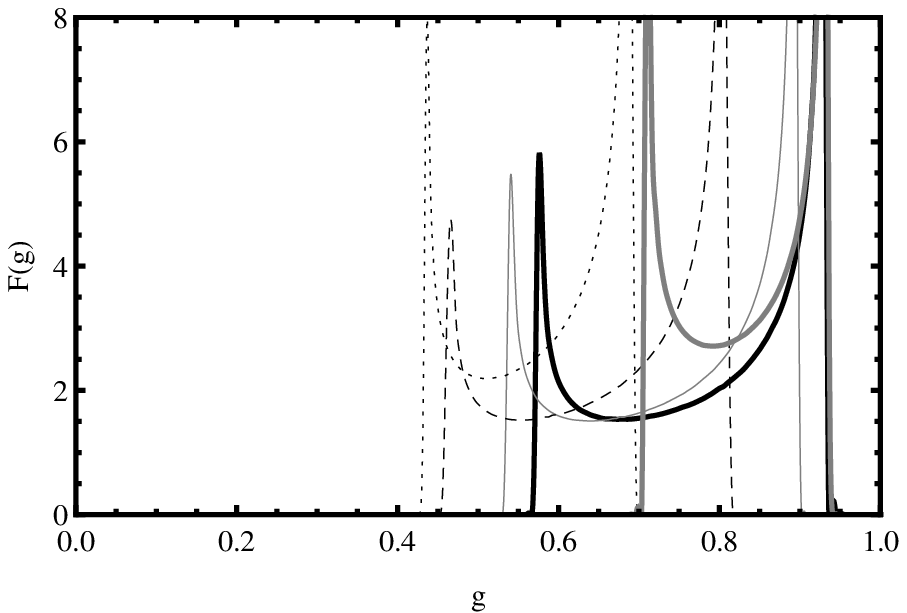}\\
\includegraphics[scale=0.7]{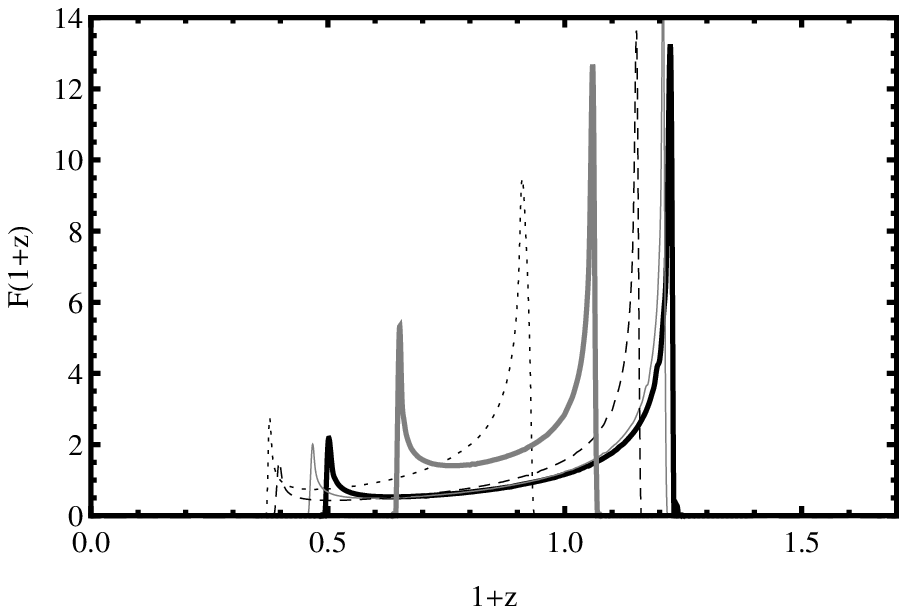}&\includegraphics[scale=0.7]{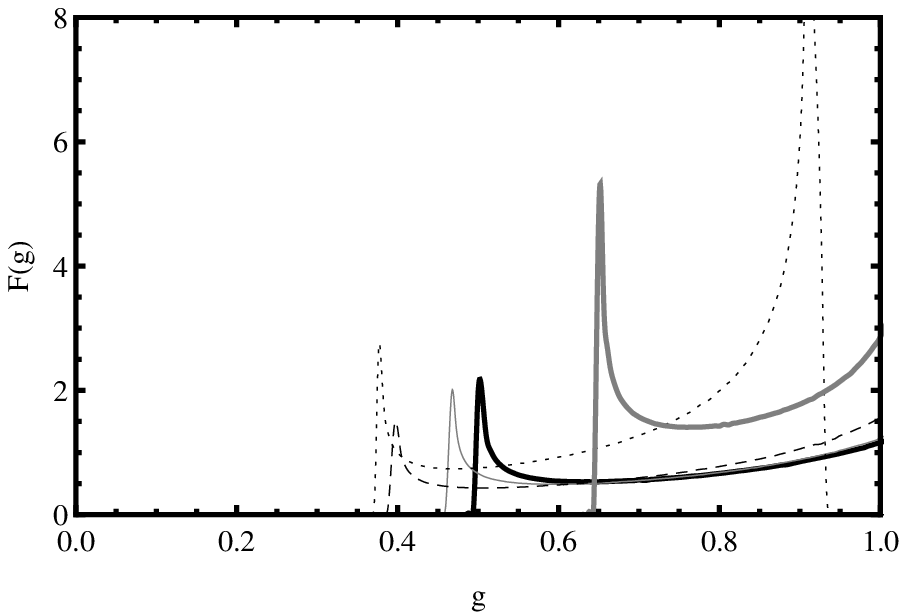}\\
\includegraphics[scale=0.7]{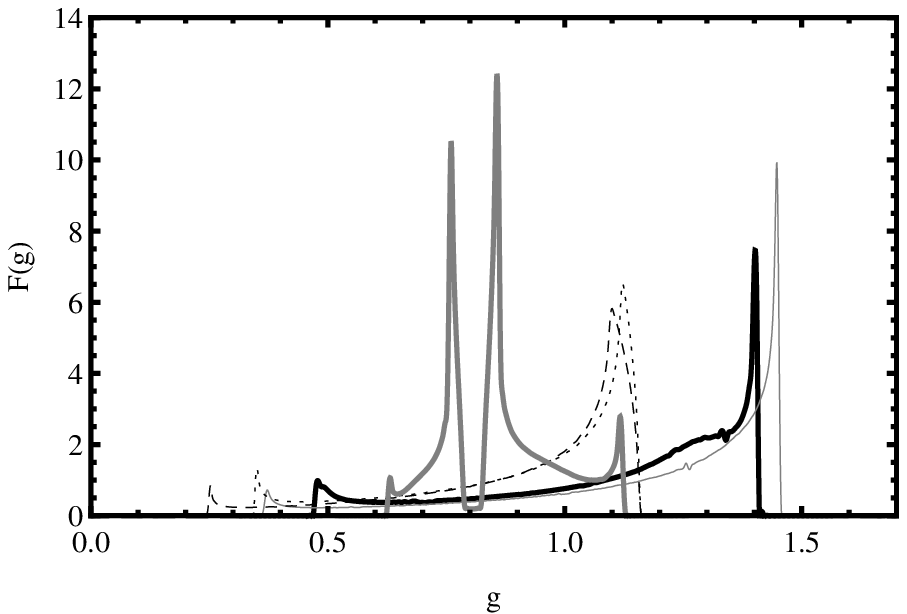}&\includegraphics[scale=0.7]{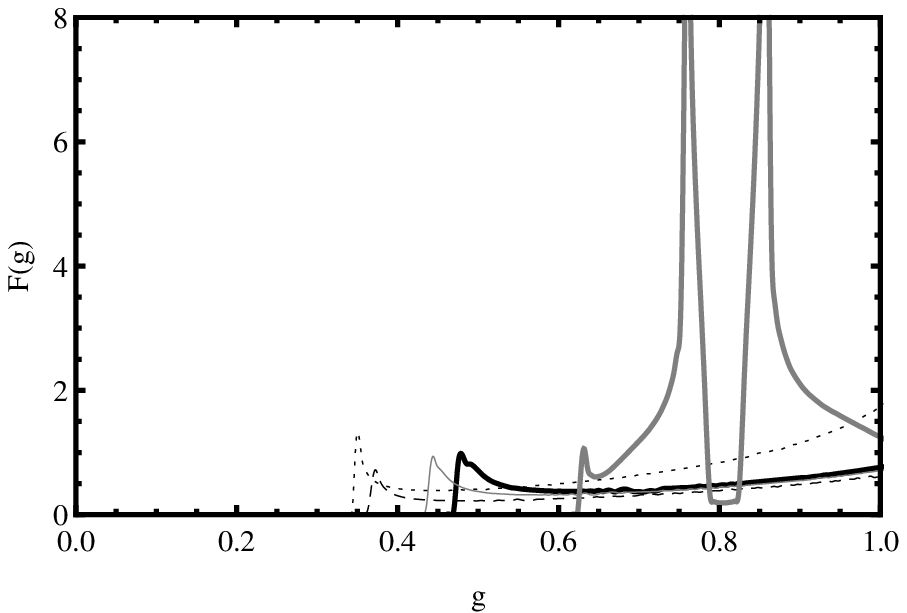}
\end{tabular}
\caption{Profiles of the spectral lines generated by Keplerian rings in the \emph{ABG} spacetimes with parameter 
$g = 0.5$ (solid,gray), $0.65$ (dashed,black), $0.72$ (dotted,black), and $1.5$ (thick,gray) and compared to those generated in the Schwarzschild spacetime (thick, black). The profiles were constructed for the three representative values of observer inclination $\theta_o = 30^\circ$ (top), $\theta_o = 60^\circ$ (middle), and $\theta_o = 85^\circ$ (bottom). Location of the Keplerian ring in the \emph{ABG} spacetime is at $r_{ISCO}$ (for $g \leq 0.9$), and at $r_{\Omega MAX}$ (for $g=1.5$), and in the Schwarzschild spacetime it is at $r_{ISCO} = 6$. In the right column, magnification of the red end of the profiled lines presented in the left column is given.\label{fig.11} }
\end{center}
\end{figure}

\begin{figure}[H]
\begin{center}
\begin{tabular}{cc}
\includegraphics[scale=0.7]{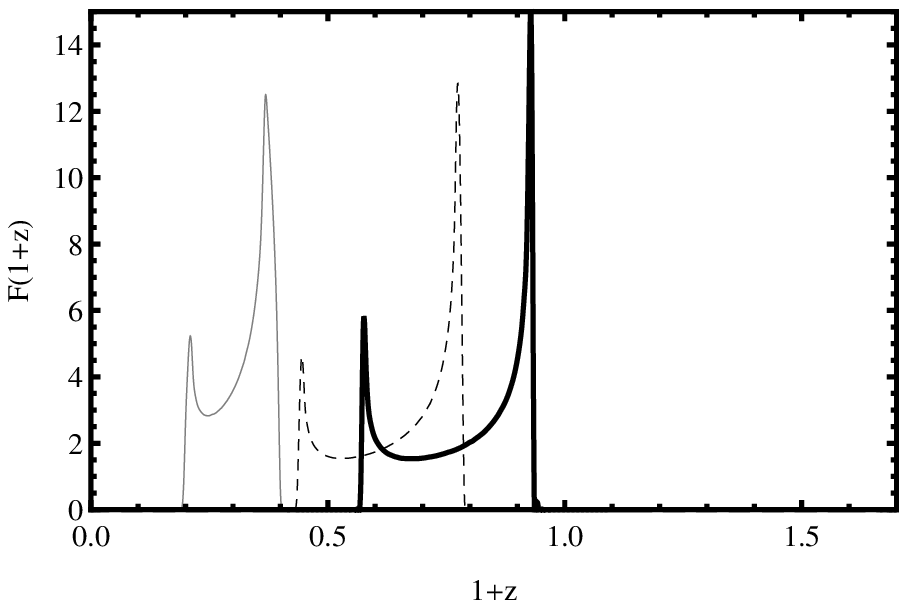}&\includegraphics[scale=0.7]{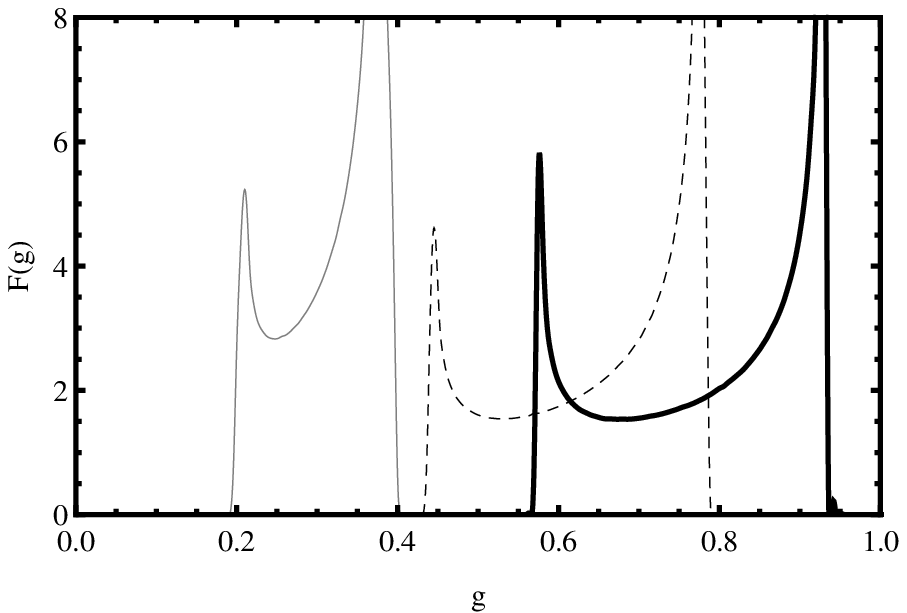}\\
\includegraphics[scale=0.7]{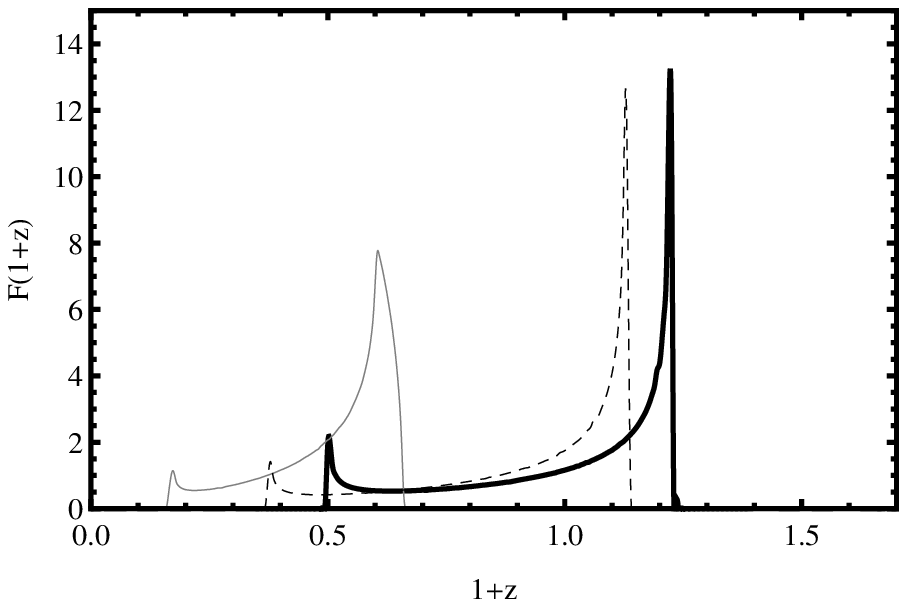}&\includegraphics[scale=0.7]{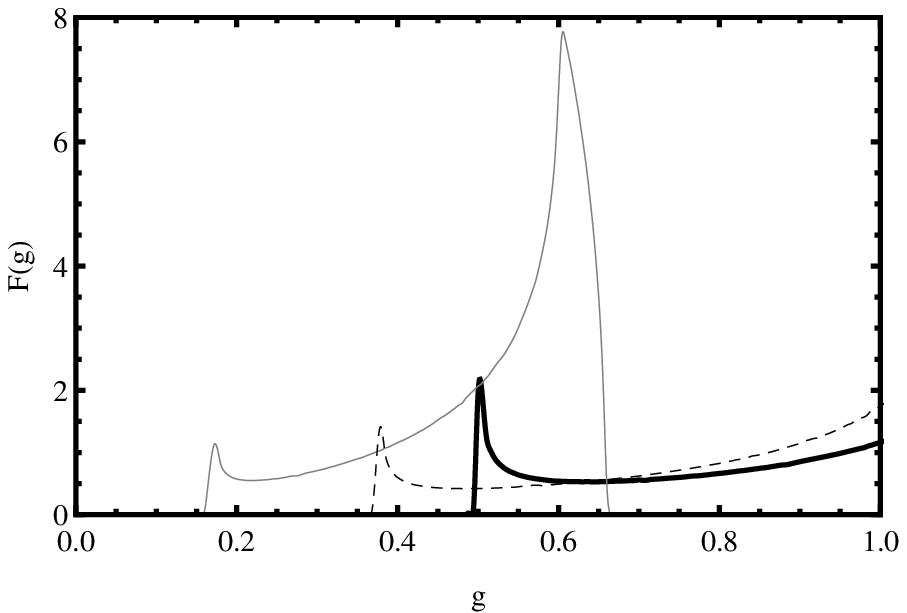}\\
\includegraphics[scale=0.7]{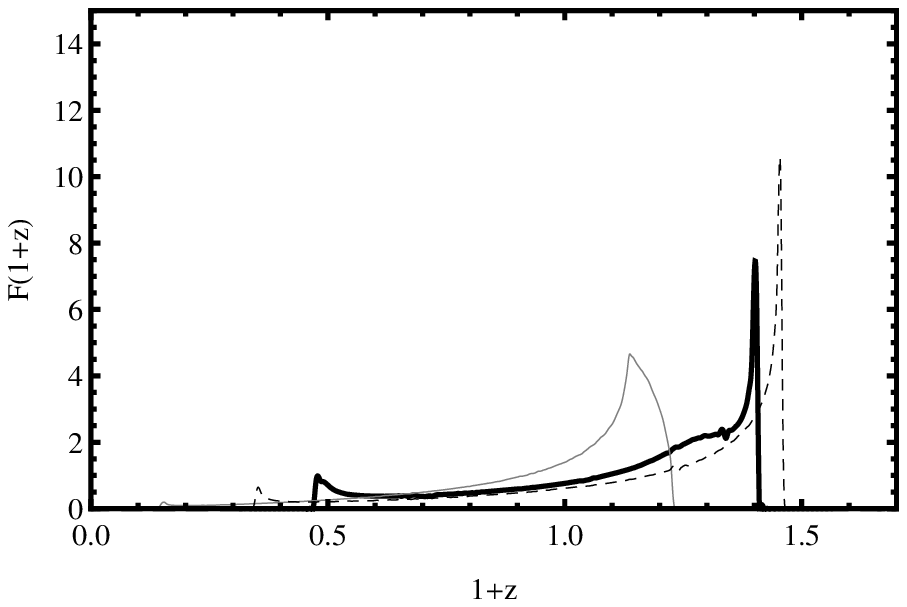}&\includegraphics[scale=0.7]{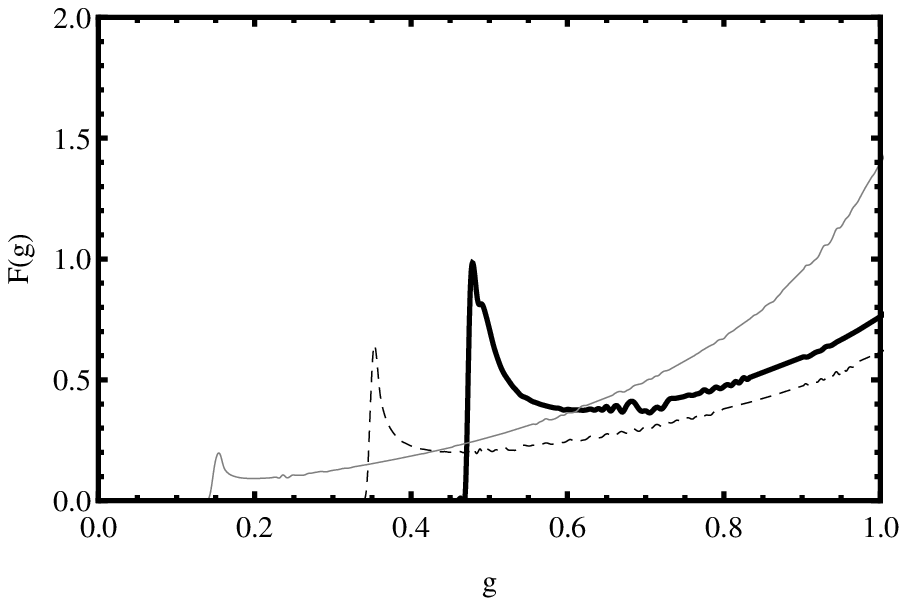}
\end{tabular}
\caption{Profiles of the spectral lines generated by Keplerian rings located in the \emph{ABG} spacetime with the charge parameter $g=0.72$ at the OSCO of the inner region of the stable circular geodesics (gray) and at the ISCO of the outer region of stable circular geodesics (dashed). They are compared to the line generated at the ISCO in the Schwarzschild spacetime (thick, black). The profiles were constructed for the three representative values of observer inclination $\theta_o=30^\circ$(top), $\theta_o=60^\circ$ (middle), and $\theta_o=85^\circ$ (bottom).\label{fig.12}  }
\end{center}
\end{figure}

\begin{figure}[H]
\begin{center}
\begin{tabular}{ccc}
	\includegraphics[scale=0.5]{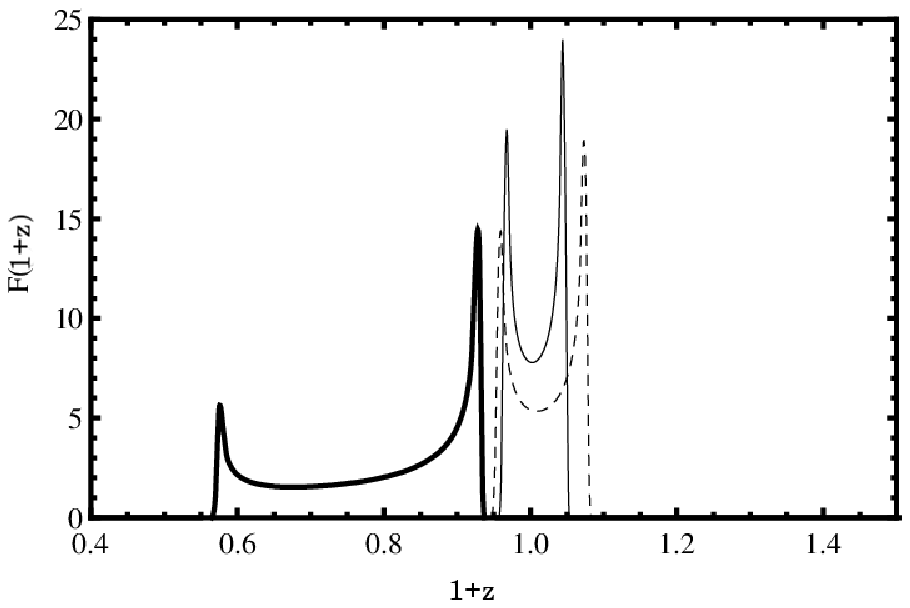}&\includegraphics[scale=0.5]{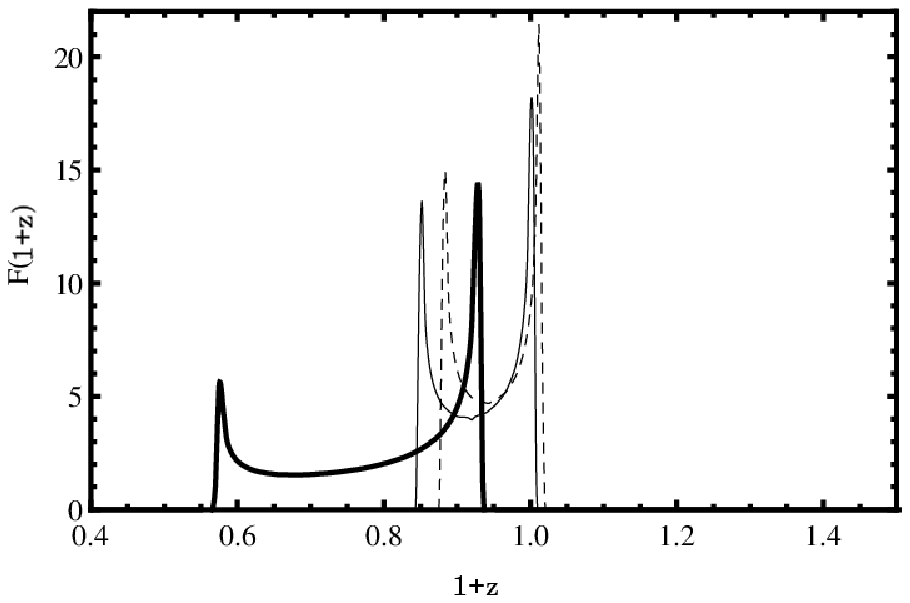}&\includegraphics[scale=0.5]{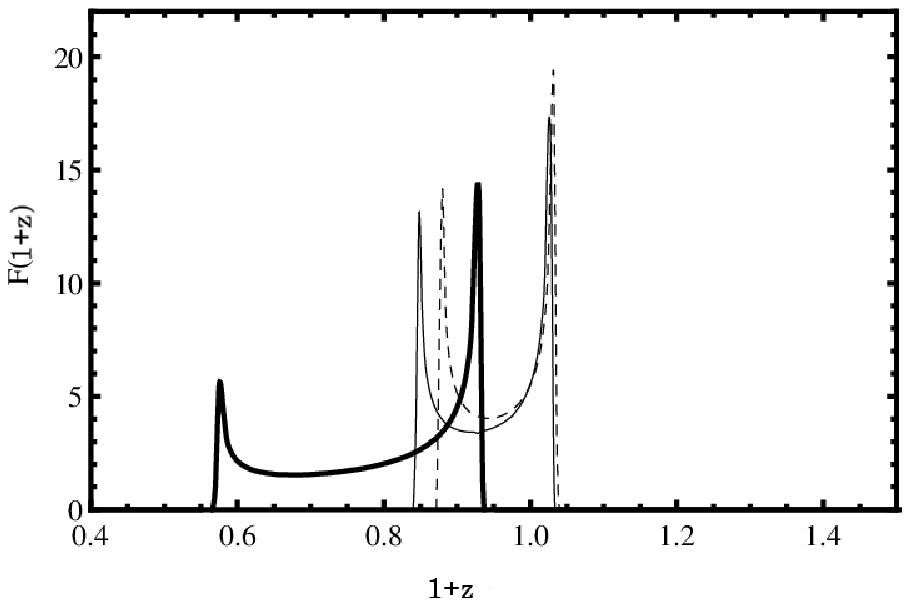}\\
	\includegraphics[scale=0.5]{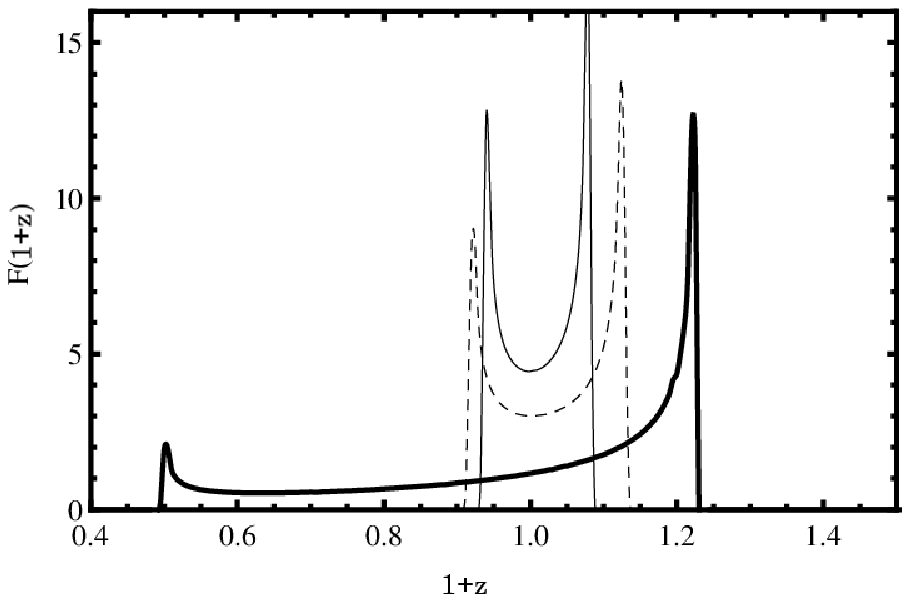}&\includegraphics[scale=0.5]{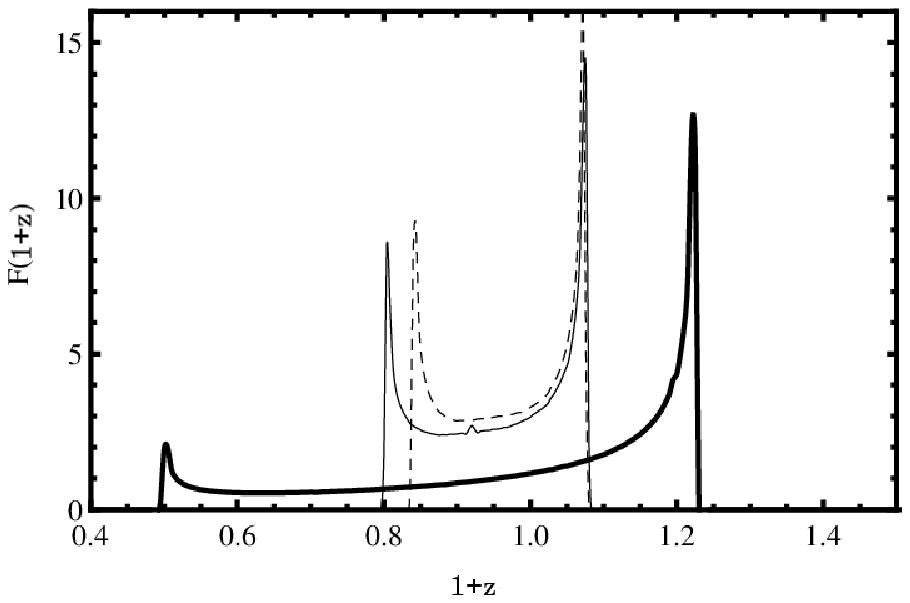}&\includegraphics[scale=0.5]{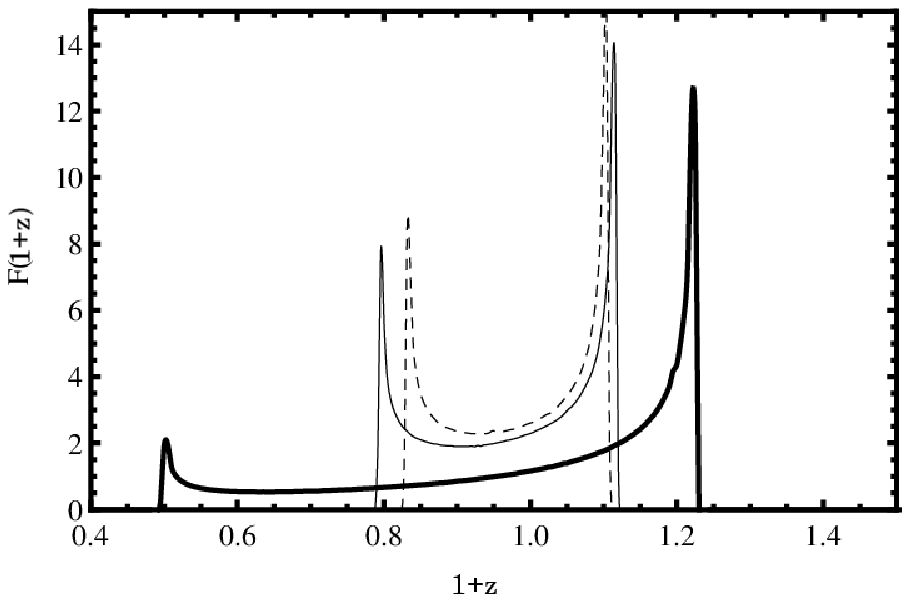}\\
	\includegraphics[scale=0.5]{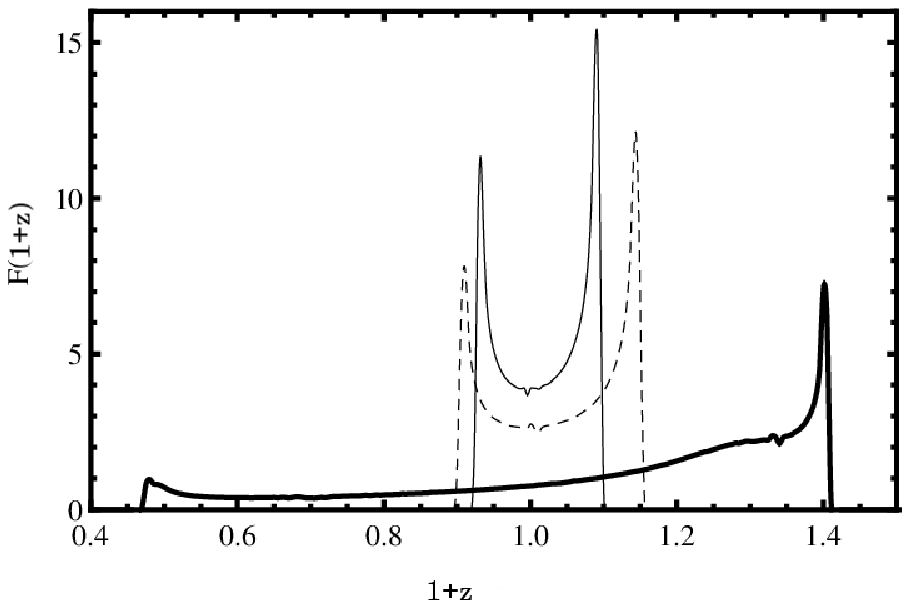}&\includegraphics[scale=0.5]{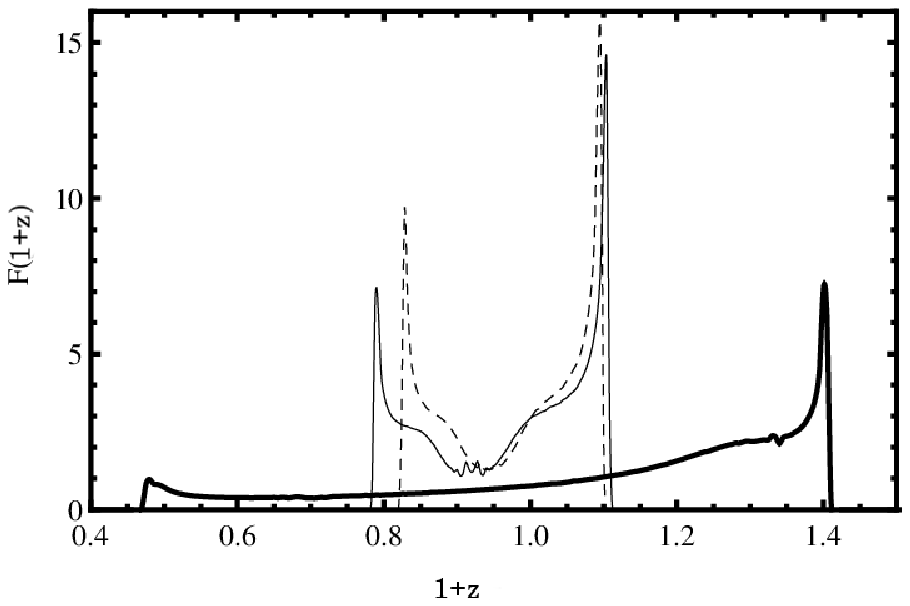}&\includegraphics[scale=0.5]{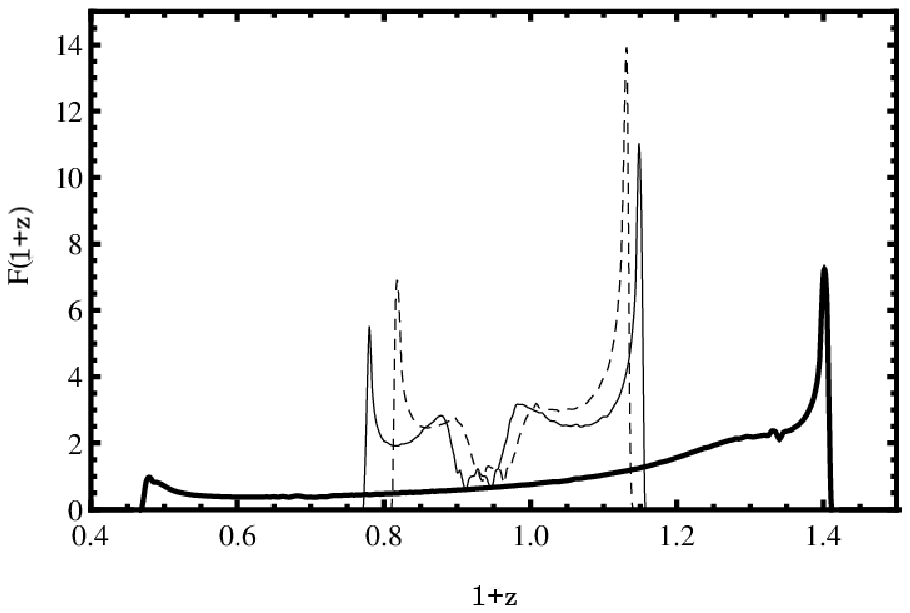}
\end{tabular}
\caption{Profiles of the spectral lines generated by Keplerian rings in the \emph{ABG} spacetimes with the charge parameter $g=2.5$ (solid) and $g=3.0$ (dashed), compared to those generated in the Schwarzschild spacetime (thick). The profiles were constructed for the three representative values of observer inclination $\theta_o=30^\circ$(top), $\theta_o=60^\circ$ (middle), and $\theta_o=85^\circ$ (bottom). Location of the Keplerian ring in the \emph{ABG} spacetime is at $r_{ISCO}$ (left), $r_{\Omega MAX}$ (center), and $1.5\times r_{\Omega MAX}$; in the Schwarzschild spacetime it is at $r_{ISCO}=6$.\label{fig.13} }
\end{center}
\end{figure}

\section{Discussion and conclusions}

We have studied motion of test particles and photons in the regular Bardeen and ABG spacetimes of both the black hole and no-horizon type that are governed by charge parameter $g$ and treated in the framework of a non-linear electrodynamics. No apriori given restrictions on the value of the $g$ parameter were considered. 
We have focused our attention on the geodesic structure of the spacetimes, especially on the circular geodesic motion that governs Keplerian accretion discs and is relevant also for toroidal perfect fluid structures governed by an interplay of gravity and pressure gradients. Simple optical phenomena have been modelled to reflect the basical properties of the geodesic structure of the spacetimes. The silhouette of the central part of the spacetime and the profiled spectral lines of geodesic (Keplerian) rings orbiting in the strong gravity regions of the Bardeen and ABG spacetimes have been constructed to give an observationally relevant signatures of the spacetimes. 

We have demonstrated that in the case of the Bardeen and ABG black hole spacetimes, the circular orbits are of the same character as in the Schwarzschild spacetime above the outer horizon, demonstrating no qualitatively different phenomena related to astrophysical processes. No circular geodesics are allowed under the inner horizon of the Bardeen and ABG black holes, similarly to the case of the Kehagias-Sfetsos black holes \cite{Vie-etal:2014:PHYSR4:}. In both the Bardeen and ABG black hole spacetimes only small quantitative differences to the Schwarzschild spacetime have been found for both the silhouette extension, and the shape and extension of the profiled spectral lines. 

Fundamental differences in comparison to the Schwarzschild spacetime have been demonstrated for the no-horizon regular spacetimes. For both the Bardeen and ABG spacetimes the differences are in accord with the effects found in the case of the Kehagias-Sfetsos naked singularity spacetimes of the modified Ho\v{r}ava quantum gravity discussed in \cite{Vie-etal:2014:PHYSR4:,Stu-Sche-Abd:2014:PHYSR4:,Stu-Sche:2014:CLAQG:}, or the RN naked singularity spacetimes \cite{Stu-Hle:2002:ActaPhysSlov:,Pug-etal:2011:PHYSR4:}. 

For no-horizon regular Bardeen and ABG spacetimes, three different regimes of the circular geodesic motion occur above the so called antigravity sphere corresponding to a stable equilibrium of static particles in dependence on the parameter $g$. For sufficiently large values of the dimensionless charge parameter $g > g_{S}$, only stable circular geodesics exist everywhere above the static radius where the antigravity sphere is located. For smaller values of the charge parameter, $g_{P} < g < g_{S}$, two distinct regions of the stable circular geodesics occur -- the inner one begins at the static radius and terminates at an OSCO, while the outer one begins at an ISCO; these two regions are separated by a region of unstable circular geodesics. For charge parameters close to the black hole states, $g_{NoH} < g < g_{P}$, an inner region of stable circular geodesics is terminated by a stable circular photon orbit, and an outer region of the stable circular geodesics begins at an ISCO. Region of unstable circular geodesics under ISCO is terminated by an unstable photon circular geodesic. No circular geodesics are allowed between the inner stable photon orbit and the outer unstable photon orbit. 

However, in contrast to the case of the Kehagias-Sfetsos naked singularity spacetimes, a special kind of behavior of the circular geodesics has been found for the ABG no-horizon spacetimes with $g > 2$. An inner region of circular geodesics can exist under the antigravity sphere of particles in a stable equilibrium at the static radius. The region consists of both stable and unstable circular geodesics, the stable geodesics are limited by $r=0$ and the OSCO, while the unstable geodesics are limited by a secondary static radius corresponding to static particles (having $L=0$) being in an unstable equilibrium position. This interesting change of the character of circular geodesics in the ABG no-horizon spacetimes with $g > 2$ is probably connected to the dramatic change of the ABG spacetime structure near the origin of coordinates where the lapse function changes its character from the "de Sitter" form corresponding to a vacuum energy implying gravitational repulsion to the "anti-de Sitter" form related to gravitational attraction. Note that no such transition occurs in the Bardeen spacetimes where the lapse function has always the "de Sitter" form near the origin of coordinates. Similar gravitational repulsion occurs in the lapse function of the KS spacetimes near the coordinate origin, however, in this case the gravitational repulsion is not of the "de Sitter" type, but corresponds rather to a repulsive effect of a quintessential type \cite{Stu-Sche:2014:CLAQG:}. 

The astrophysically most profound property of the Bardeen and ABG no-horizon spacetimes is existence of the antigravity sphere -- however, the same phenomenon is expected in the case of the naked singularity spacetimes of the RN or Kehagias-Sfetsos type. The observationally relevant consequences of the existence of the Bardeen or ABG no-horizon spacetimes are related to the profiled spectral lines generated by a Keplerian ring radiating at a given frequency in the strong gravity regions of these spacetimes. Our results clearly demonstrate qualitative and strong quantitative differences in comparison to profiled lines generated in the Schwarzschild spacetime. Moreover, there are also strong differences when the effects on the profiled spectral lines are compared to those related to the Kehagias-Sfetsos naked singularities. In addition, the differences enable to distinguish easily also the Bardeen and ABG no-horizon spacetimes, if inclination angle of the observer is known. 

We can conclude that the no-horizon Bardeen and ABG spacetimes could give very strong signatures in the observationally relevant phenomena related to the profiled spectral lines that enable to distinguish them not only from the black hole spacetimes, but also from the Kehagias-Sfetsos naked singularity spacetimes. Therefore, there are some observationally relevant effect enabling to obtain distinct predictions of the Einstein gravity combined with non-linear electrodynamics, and the modified Ho\v{r}ava quantum gravity. 

\section*{Acknowledements}
Z.S. and J.S. acknowledge the Albert Einstein Centre for Gravitation and Astrophysics supported by the~Czech Science Foundation Grant No. 14-37086G.

\end{document}